\documentclass[aps,twocolumn,superscriptaddress]{revtex4-1} 
\usepackage{graphicx}
\usepackage{amsmath,amssymb}
\usepackage{dsfont}
\usepackage{bm}
\usepackage{MnSymbol, marvosym, wasysym}
\usepackage{float}
\usepackage{color}
\usepackage{soul}

\renewcommand{\vec}[1]{\boldsymbol{#1}}

\newcommand{\pd}{{\phantom{\dagger}}}
\newcommand{\bs}[1]{\boldsymbol{#1}}

\newcommand{\ie}{{\it i.e.},\ }

\graphicspath{{./}{./figures/}}

\begin{document}

\title{Competition of first-order and second-order topology on the honeycomb lattice}

\author{Matthew Bunney}
\affiliation{School of Physics, University of Melbourne, Parkville, VIC 3010, Australia}
\author{Tomonari Mizoguchi}
\affiliation{Department of Physics, University of Tsukuba, Tsukuba, Ibaraki 305-8571, Japan}
\author{Yasuhiro Hatsugai}
\affiliation{Department of Physics, University of Tsukuba, Tsukuba, Ibaraki 305-8571, Japan}
\author{Stephan Rachel}
\affiliation{School of Physics, University of Melbourne, Parkville, VIC 3010, Australia}

\date{\today}

%%%%%%%%%%%%%%%%%%%%%%%%%%%%%%%%%%%%%%%%%%%%%%%%%%%%%%%%%%%%%%%%%%%%%%%%%%%

\begin{abstract}
We investigate both first-order topology, as realized through Haldane's model, and second-order topology, implemented through an additional Kekul\'e-distortion, on the honeycomb lattice. The interplay and competition of both terms result in a phase diagram at half-filling which contains twelve distinct phases. All phases can be characterized by the first Chern number or by a quantized $\mathbb{Z}_Q$ Berry phase. Highlights include phases with high Chern numbers, a novel $\mathbb{Z}_6$ topological phase, but also coupled kagome-lattice Chern insulators. Furthermore, we explore the insulating phases at lower fillings, and find again first-order and second-order topological phases. Finally, we identify real-space structures which feature corner states not only at half but also at third and sixth fillings, in agreement with the quantized  $\mathbb{Z}_Q$ Berry phases.
\end{abstract}

\maketitle

%%%%%%%%%%%%%%%%%%%%%%%%%%%%%%%%%%%%%%%%%%%%%%%%%%%%%%%%%%%%%%%%%%%%%%%%%%%
%%%%%%%%%%%%%%%%%%%%%%%%%%%%%%%%%%%%%%%%%%%%%%%%%%%%%%%%%%%%%%%%%%%%%%%%%%%

%%%%%%%%%%%%%%%%%%%%%%%%%%%%%%%%%%%%%%%%%%%%%%%%%%%%%%%%%%%%%%%%%%%%%%%%%%%
%
%                                                                            I N T R O D U C T I O N
%
%%%%%%%%%%%%%%%%%%%%%%%%%%%%%%%%%%%%%%%%%%%%%%%%%%%%%%%%%%%%%%%%%%%%%%%%%%%

%\textcolor{red}{Red: additions by TM}\\
\section{Introduction}
\label{sec:intro}

Topological insulators (TIs) are bulk insulators with topologically protected metallic states confined to the boundary\,\cite{hasan2010,qi2011,bernevig2013}, immune to single-particle backscattering\,\cite{halperin82prb2185,hatsugai93prb11851,PhysRevB.73.045322}. They are a generalisation of the integer quantum Hall effect\,\cite{klitzing-80prl494}, which is characterized by the Chern number\cite{thouless-82prl405}. More specifically, these topological insulators are fermionic phases protected by time-reversal symmetry and $U(1)$ charge conservation. Today, we know a plethora of such symmetry-protected topological (SPT) phases\,\cite{Volovik03,pollmann-10prb064439,chen-11prb235128,chen2012,wen2012}. In particular, it has been claimed that the one-dimensional fermionic SPT phases have been fully classified\,\cite{pollmann-10prb064439,chen-11prb235128}. Moreover, all non-interacting fermionic phases in $d$ dimensions have been characterized based on both group cohomology  and $K$-theory\,\cite{schnyder2008,kitaev2009,wen2012}.
SPT phases usually have boundary (\ie edge or surface) modes present and their bulk states can be classified by a topological invariant, a phenomenon referred to as {\it bulk boundary correspondence}\,\cite{hatsugai93prl3697,ryu-02prl077002}.

The time-reversal invariant TIs in two dimensions (2D) represent the pioneering subject in this young field, also known as quantum spin Hall (QSH) insulators\,\cite{kane2005,kane2005a,bernevig2006a}. The proposal that the QSH effect might exist in negative band gap semiconductor wells\,\cite{bernevig2006} was swiftly followed with the first experimental discovery of a 2D TI in HgTe/CdTe quantum wells\,\cite{konig2007}. The other early proposal for 2D QSH effect was for the honeycomb lattice material graphene\,\cite{kane2005,kane2005a}; the seminal papers of Kane and Mele also highlighted that a spinless, time-reversal breaking version of their honeycomb model was introduced by Haldane 16 years earlier\,\cite{haldane1988}. 

The electric multipole insulators in the work of Benalcazar, Bernevig and Hughes (BBH)\,\cite{benalcazar2017, benalcazar2017a} generalize bulk boundary correspondence beyond the previously mentioned models of TIs. While those TIs in $d$ dimensions exhibit $d-1$ dimensional boundary states, these multipole insulators feature $d-n$ dimensional corner or hinge states. These modes correspond to quantized higher electric multipole moments, and because of this unusual bulk-boundary correspondence they are often referred to as {\it higher-order} TIs (HOTIs)\,\cite{schindler2018, schindler2018a, langbehn2017, song2017}, in contrast to the previously mentioned first-order TIs. HOTIs represent today an active and flourishing field of research\,\cite{ezawa2018-kagome,ezawa2018, elcoro2017, hashimoto2017, hayashi2018, wakao2020, takahashi2021, watanabe2020, mizoguchi2020, trifunovic2019, khalaf2018}. Their theoretical prediction has been quickly matched with the first experimental realizations, with HOTIs realized in bismuth\,\cite{schindler2018,schindler2018a}, topolectrical circuits\,\cite{imhof2018, kempkes2019, zangeneh-nejad2019, wieder2021, yang2020, fan2019}, 
photonic crystals\,\cite{noh2018,ota2019,elhassan2019},
acoustic\,\cite{xue2019,Xue2019_acoustic,Ni2019} and elastic systems\,\cite{fan2019}. Usually one considers first- and higher-order topology as exclusive: either one or the other is realized. That might be the reason why only little work has been done on the combined effects of first- and higher-order topology. 

The honeycomb lattice offers some of the simplest pathways to modelling topologically non-trivial behaviour. For example, the Dirac bandstructure can be realized on the honeycomb lattice without the need of any $\pi$ fluxes, while the square lattice with its trivial parabolic bands switches to Dirac bands only when subject to $\pi$ flux per plaquette. In a similar fashion, the simplest HOTI model, the square lattice BBH model\,\cite{benalcazar2017, benalcazar2017a} , still requires insertion of $\pi$ fluxes, which are not necessary for the honeycomb lattice. The honeycomb lattice offers second-order topology solely through anisotropic hopping amplitudes, referred to as Kekul\'e and anti-Kekul\'e distortions\,\cite{ajiki-jpsj260,chamon00prb2806,hou-07prl186809,frank-11prl066801} which was recently realized in graphene\,\cite{watanabe2020}. 
Several works have explored HOTI phases on the honeycomb lattice\,\cite{sato2008, gomes2012, guinea2010, hasegawa2012, lee2020, liu2019, mizoguchi2019, wu2016, radha2020,diop2020} including experimental work\,\cite{zangeneh-nejad2019}.
The simple Kekul\'e distortion and the connection to Kane and Mele's QSH model (for spinful electrons) or to Haldane's Chern insulator model (for spinless electrons), respectively, both make the honeycomb lattice the perfect testing ground to explore the interplay and competition of first- and second-order topology. The Kane-Mele model with Kekul\'e anisotropy was previously studied in Ref.\,\cite{wu-12prb205102} and more recently an experimentally motivated variant of Haldane's model with Kekul\'e distortions\,\cite{li2020}.

In this paper, we study the interplay and competition of first- and second-order topology of spinless electrons. In Sec.\,\ref{sec:setup}, we introduce the model of our investigation, a Kekul\'e--distorted Haldane model. In Sec.\,\ref{sec:Chern-phases}, we analyze the Chern phases at half filling, and in Sec.\,\ref{sec:hoti-phases} we analyze HOTI phases at half filling. Sec.\,\ref{sec:lower-fillings} contains our analysis of the phases that emerge at lower fillings. After a discussion of the results in Sec.\,\ref{sec:discussion}, the paper ends with a summary in Sec.\,\ref{sec:summ}.

%%%%%%%%%%%%%%%%%%%%%%%%%%%%%%%%%%%%%%%%%%%%%%%%%%%%%%%%%%%%%%%%%%%%%%%%%%%
%%%%%%%%%%%%%%%%%%%%%%%%%%%%%%%%%%%%%%%%%%%%%%%%%%%%%%%%%%%%%%%%%%%%%%%%%%%
%%%%%%%%%%%%%%%%%%%%%%%%%%%%%%%%%%%%%%%%%%%%%%%%%%%%%%%%%%%%%%%%%%%%%%%%%%%

%%%%%%%%%%%%%%%%%%%%%%%%%%%%%%%%%%%%%%%%%%%%%%%%%%%%%%%%%%%%%%%%%%%%%%%%%%%
%
%                                                           M O D E L    A N D    D E F I N I T I O N S
%
%%%%%%%%%%%%%%%%%%%%%%%%%%%%%%%%%%%%%%%%%%%%%%%%%%%%%%%%%%%%%%%%%%%%%%%%%%%
\section{Model and Definitions}
\label{sec:setup}
\begin{figure}[t!]
\centering
\includegraphics[width = 0.39\textwidth]{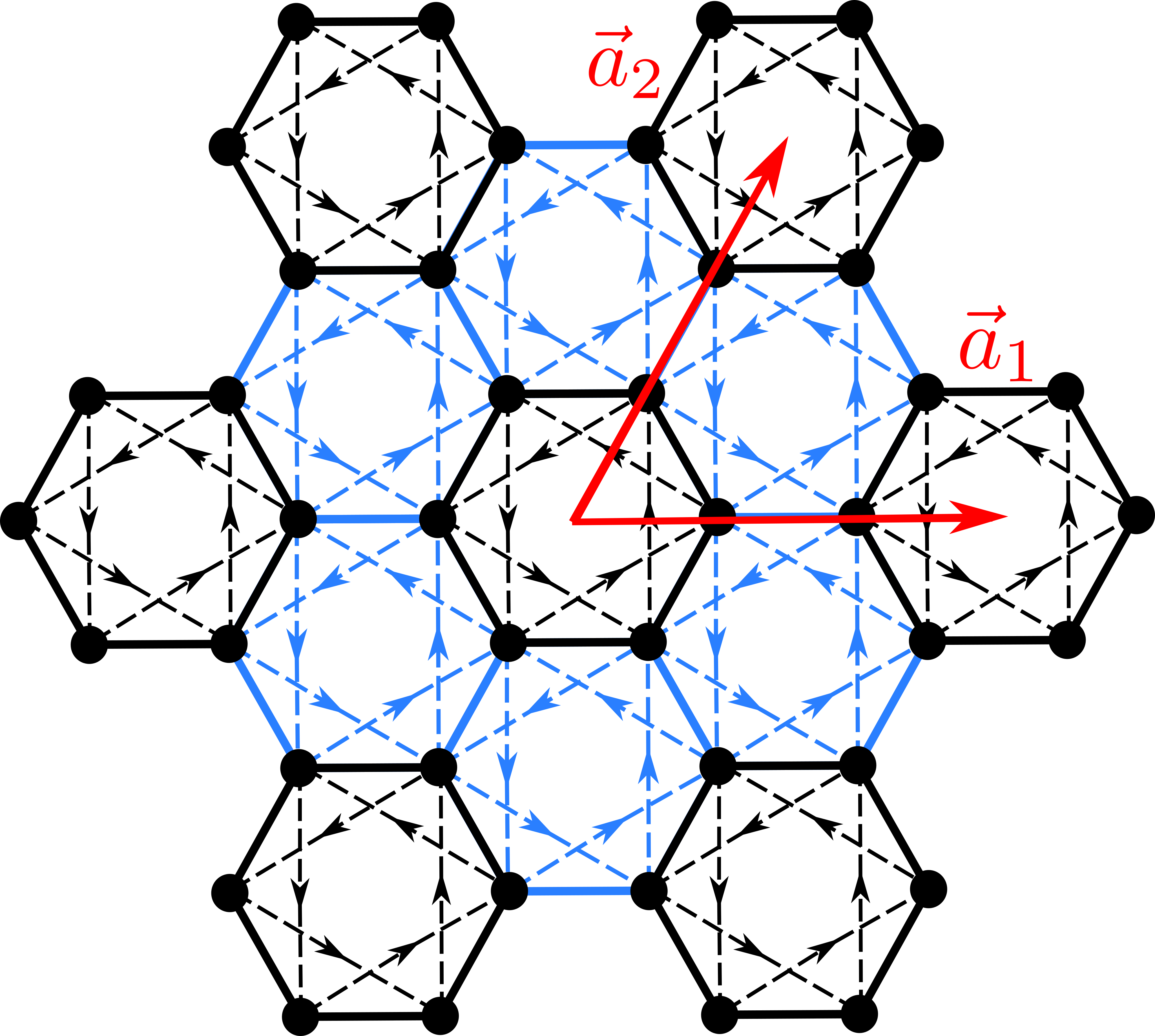}
\caption{Honeycomb lattice: nearest-neighbor bonds shown as solid lines, second-neighbor bonds as dashed lines, and the arrows fix the phase convention of the Haldane term. The ratio of hoppings of the blue and black bonds is given by $\alpha$, quantifying the Kekul\'e distortion. Lattice vectors $\bs{a}_1$ and $\bs{a}_2$ are shown in red.}
\label{fig:hoti_lattice}
\end{figure}

The Hamiltonian, containing both Haldane's term and a Kekul\'e distortion on the honeycomb lattice, is given by
\begin{equation}
\begin{split}
	H &=  s \left(t \sum_{\langle i, j \rangle} c_{i}^{\dagger} c_{j}^\pd  + t' \sum_{\langle \!\langle i, j \rangle \!\rangle} e^{\pm i \phi} c_{i}^{\dagger} c_{j}^\pd \right) \quad\!\!\text{{\color{black}(hexamers)}} \\
	&+ s'\left( t \sum_{\langle i, j \rangle} c_{i}^{\dagger} c_{j}^\pd + t' \sum_{\langle \!\langle i, j \rangle \!\rangle} e^{\pm i \phi} c_{i}^{\dagger} c_{j}^\pd \right) \quad\!\!\text{{\color{blue}(between hexamers)}} 
\end{split}
\label{ham}
\end{equation}

where $\langle i,j \rangle$ ($\langle \! \langle i,j \rangle \!\rangle$) denotes (next) nearest neighbour lattice sites.
Thus $t$ ($t'$) is the real-valued hopping amplitude of the (second-) nearest neighbour bonds. In the $t'$ term, the phase factor has a positive (negative) exponent for second-nearest neighbour hoppings that go counter-clockwise (clockwise) around a hexamer. We set $\phi = \pi/2$ which makes the second-neighbor hopping purely imaginary, breaking time-reversal symmetry explicitly. That is the essence of Haldane's term and can enable anomalous quantum Hall effect. To introduce the Kekul\'e or anti-Kekul\'e distortion, respectively, the Hamiltonian is split into two parts. The first part contains all bonds within the hexamers (shown in black in Fig.\,\ref{fig:hoti_lattice}), modulated by an overall scaling factor $s$. The second part contains all other bonds, \ie all bonds between hexamers (shown in blue in Fig.\,\ref{fig:hoti_lattice}), with scaling factor $s'$. 

The unit cell for this model contains six atoms (a hexamer), spanned by the following lattice vectors:
\begin{equation}
	\vec{a}_1 = 3a \vec{\hat{x}}, \quad \vec{a}_2 = \frac{3a}{2} ( \vec{\hat{x}} + \sqrt{3} \vec{\hat{y}})\ .
\end{equation}
We further introduce $\vec{a}_3 = \vec{a}_1 - \vec{a}_2$; $a$ is the lattice spacing, and $\vec{\hat{x}}$ and $\vec{\hat{y}}$ are unit vectors in the $x$ and $y$-directions, respectively. The reciprocal lattice vectors are then given as
\begin{equation}
	\vec{b}_1 =\frac{2 \pi}{3 \sqrt{3} a} ( \sqrt{3} \vec{\hat{x}} - \vec{\hat{y}}), \quad \vec{b}_2 = \frac{4 \pi}{3 \sqrt{3} a} \vec{\hat{y}}\ .
\end{equation}

For convenience, we set $a=1$ throughout the paper. By imposing periodic boundary conditions, we can express  the model \eqref{ham} in momentum space,
\begin{equation}
	\label{eq:hamiltonian_k}
	H = \sum_{\vec{k}} \vec{c}_{\vec{k}}^\dagger h (\vec{k}) \vec{c}^\pd_{\vec{k}}\ ,
\end{equation}
where $\vec{c}_{\vec{k}}$ is the vector of annihilation operators $(c_{1,\vec{k}}, ..., c_{6,\vec{k}})^T$, and $h(\vec{k})$ is the Bloch matrix.
\begin{align}
	\label{eq:bloch_hamiltonian}
	&h = it'
	\begin{pmatrix}
		0 & 0 & A & 0 & -B^* & 0 \\
		0 & 0 & 0 & B^* & 0 & -C^* \\
		-A^* & 0 & 0 & 0 & C & 0 \\
		0 & -B & 0 & 0 & 0 & A^* \\
		B & 0 & -C^* & 0 & 0 & 0 \\
		0 & C^* & 0 & -A & 0 & 0 \\
	\end{pmatrix} \nonumber \\[5pt]
	+& t \begin{pmatrix}
		0 & s & 0 & s' e^{ -i \vec{a}_2 \cdot \vec{k}} & 0 & s\\
		s & 0 & s & 0 & s' e^{i \vec{a}_3 \cdot \vec{k} } & 0 \\
		0 & s & 0 & s & 0 & s' e^{i \vec{a}_1 \cdot \vec{k}} \\
		s' e^{i \vec{a}_2  \cdot\vec{k}} & 0 & s & 0 & s & 0 \\
		0 & s' e^{-i \vec{a}_3 \cdot \vec{k}} & 0 & s & 0 & s\\
		s & 0 & s' e^{- i \vec{a}_1 \cdot \vec{k}} & 0 & s & 0 \\
	\end{pmatrix}
\end{align}
where
\begin{align}
A =&~ s + s' \left(e^{-i \vec{a}_1 \cdot \vec{k}}+e^{-i \vec{a}_2 \cdot \vec{k} }\right)\ , \\[5pt]
B =&~ s + s' \left(e^{-i \vec{a}_3 \cdot \vec{k} }+e^{i \vec{a}_2 \cdot \vec{k} }\right)\ , \\[5pt]
C =&~ s + s' \left(e^{i \vec{a}_3 \cdot \vec{k} }+ e^{i \vec{a}_1 \cdot \vec{k} }\right)\ ,
\end{align}
and $A^*$ is the complex conjugate of $A$.

%%%%%%%%%%%%%%%%%%%%%%%%%%%%%%%%%%%%%%%%%%%%%%
%%%%%%%%%%%%%%%%%%%%%%%%%%%%%%%%%%%%%%%%%%%%%%

The Hamiltonian \eqref{ham} can be fully parametrized by two values, which we define to be $\lambda$ and $\alpha$. (i) $\lambda = t' / t$ is the relative strength of the next-nearest neighbor (Haldane) bonds to the nearest neighbour bonds, and is a measure of ``first-order topology'' in the system. (ii) $\alpha = s' / s$ is the relative strength of bonds within hexamers to bonds between hexamers (for bonds of the same type), and is a measure of ``second-order topology'' in the system.

We use the following conventions:
\begin{equation}
\begin{split}
&\lambda \leq 1 ~~\Rightarrow~~ t=1 \wedge \lambda = t' \\
&\lambda  > 1 ~~\Rightarrow~~ t'=1 \wedge \lambda = 1/t
\end{split}
\end{equation}
Following this definition, $\lambda = \infty$ corresponds to $t' = 1 \wedge t = 0$. The same convention is used for $\alpha$. 
\begin{equation}
\begin{split}
&\alpha \leq 1 ~~\Rightarrow~~ s=1 \wedge \alpha = s' \\
&\alpha  > 1 ~~\Rightarrow~~ s'=1 \wedge \alpha = 1/s
\end{split}
\end{equation}
We thus avoid that for $\alpha=1/0$ or $\lambda=1/0$, respectively, the bandwidth going to infinity.

\begin{figure}[t!]
	\begin{center}
		\includegraphics{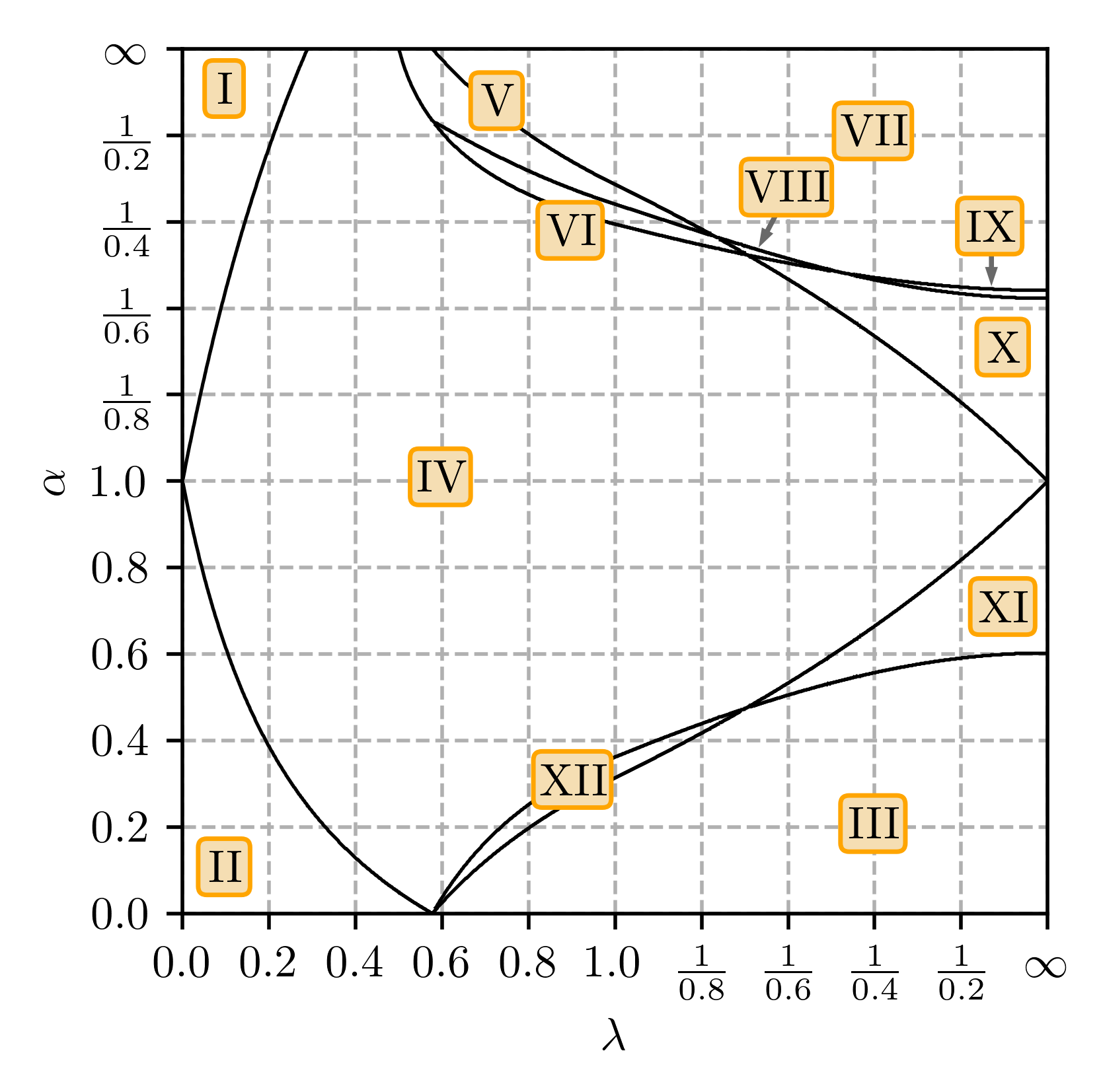}
		\caption{
			The phase boundary diagram for the combined tight-binding Hamiltonian (\ref{ham}). $\alpha$ is the ratio $s'/s$, and $\lambda$ is the ratio $t'/t$. The phase boundaries are calculated by numerically finding where the Bloch Hamiltonian (\ref{eq:bloch_hamiltonian}) has a gap closing.
		}
		\label{fig:phase_boundary_diagram}
	\end{center}
\end{figure}

Fig.\,\ref{fig:phase_boundary_diagram} shows the $\alpha$-$\lambda$ phase diagram at half filling, containing 12 different phases. The phase diagram was calculated by exact diagonalization of Eqn.\,\eqref{eq:hamiltonian_k} and not for the real-space Hamiltonian Eqn.\,\eqref{ham} in order to avoid finite-size effects. Black lines correspond to gap closings between the different phases. 

%%%%%%%%%%%%%%%%%%%%%%%%%%%%%%%%%%%%%%%%%%%%%%
%%%%%%%%%%%%%%%%%%%%%%%%%%%%%%%%%%%%%%%%%%%%%%

Since the unit cell of the system is three-fold larger than the standard two-atomic unit cell used in the tight binding model of the honeycomb lattice of graphene (which corresponds to $\alpha = 1, \lambda = 0$ in the current model), the Dirac cones located in the valleys $K$ and $K'$ in graphene's dispersion relation have been backfolded to the $\Gamma$ point, the centre of the Brillouin zone. The bulk gap closes between phases IV, I and II through this Dirac cone at $\Gamma$. The other gap closings for larger $\lambda$ do not happen at $\Gamma$, but rather at the $K$, $K'$ or $M$ and on the line between them in the new (reduced) Brillouin zone.

The bulk gap is closed for $\alpha \rightarrow \infty$, $1/2\sqrt{3} \leq \lambda \leq 1/\sqrt{3}$ (from the boundary between phases I and IV to the boundary between phases V and VII). The bulk gap also closes for $\lambda \rightarrow \infty$, which means that the bulk gap is generally of order $0.01$ or less for $\lambda \geq \frac{1}{0.2}$ (for an example, see the ribbon spectra in \ref{fig:appendix_phases_1} (e)).

%%%%%%%%%%%%%%%%%%%%%%%%%%%%%%%%%%%%%%%%%%%%%%%%%%%%%%%%%%%%%%%%%%%%%%%%%%%
%%%%%%%%%%%%%%%%%%%%%%%%%%%%%%%%%%%%%%%%%%%%%%%%%%%%%%%%%%%%%%%%%%%%%%%%%%%
%%%%%%%%%%%%%%%%%%%%%%%%%%%%%%%%%%%%%%%%%%%%%%%%%%%%%%%%%%%%%%%%%%%%%%%%%%%

%%%%%%%%%%%%%%%%%%%%%%%%%%%%%%%%%%%%%%%%%%%%%%%%%%%%%%%%%%%%%%%%%%%%%%%%%%%
%
%                                                           1 / 2 F I L L I N G   :     C H E R N    P H A S E S
%
%%%%%%%%%%%%%%%%%%%%%%%%%%%%%%%%%%%%%%%%%%%%%%%%%%%%%%%%%%%%%%%%%%%%%%%%%%%
\section{Chern Insulating Phases at half filling}
\label{sec:Chern-phases}

Nine out of the 12 phases possess a finite Chern number, are thus versions of anomalous quantum Hall effect. In the following we will discuss the most interesting of these Chern phases in detail; the other Chern phases are delegated to the Appendix.

%%%%%%%%%%%%%%%%%%%%%%%%%%%%%%%%%%%%%%%%%%%%%%
\begin{figure*}[phtb]
	\centering	
	\includegraphics[width=\textwidth]{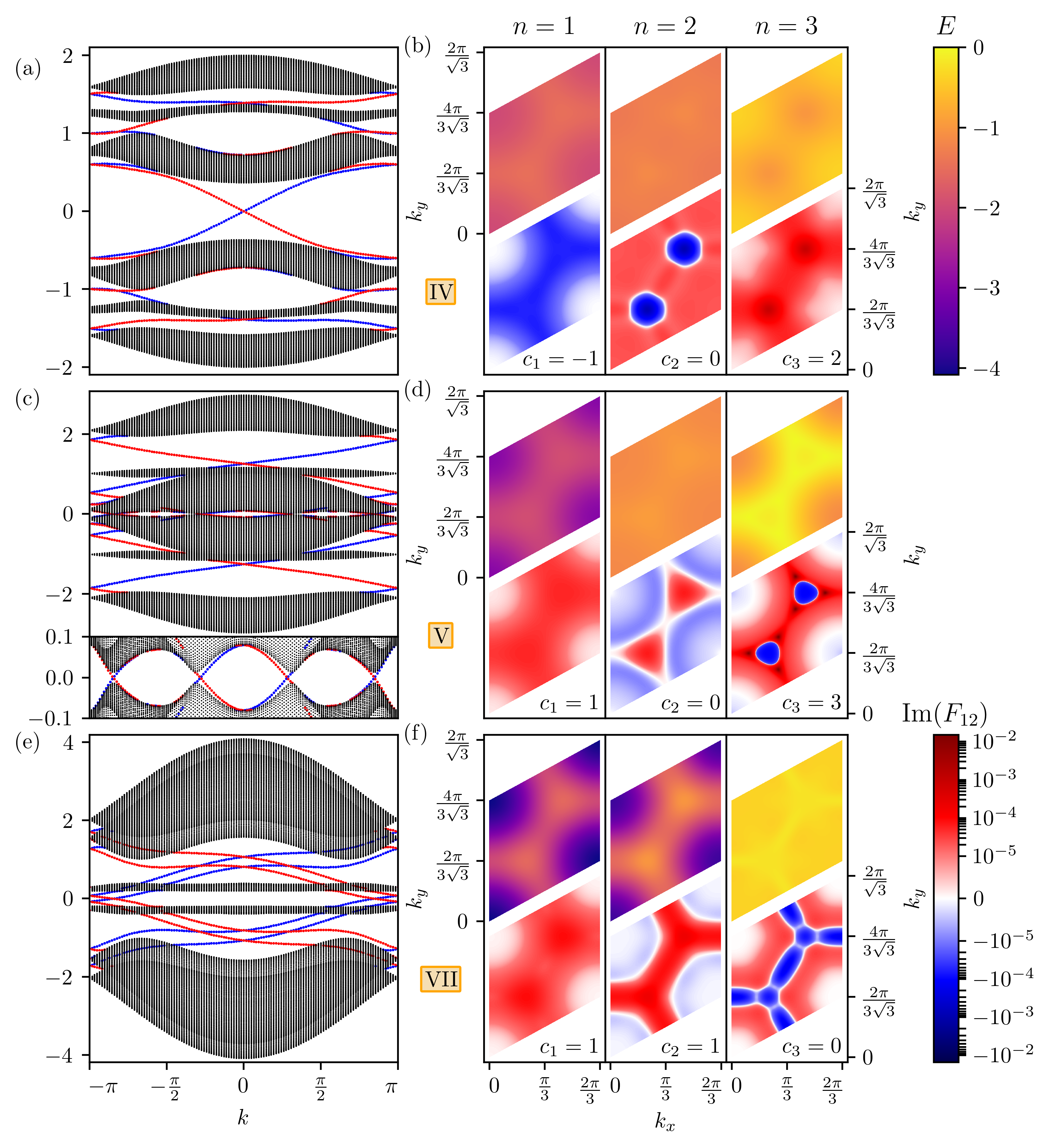}
	\caption{Ribbon spectra (a, c, e),  spectra in momentum space (top) and Berry curvature (bottom) over the Brillouin zone (b, d, f). First row (a, b) is phase IV with $C=1$ at half filling, parameters used $\alpha = \frac{1}{0.5}, \lambda = 0.2$; second row (c, d) is phase V with $C=4$ at half filling, parameters used $\alpha = \frac{1}{0.08}, \lambda = 0.57$; third row (e, f) is phase VII with $C=2$ at half filling, parameters used $\alpha = \frac{1}{0.25}, \lambda = \frac{1}{0.25}$. Displayed below Berry curvature plots are the Chern numbers $c_n$ of each individual band. For the ribbon spectra, blue (red) dots are states with $\geq$75\% of the wavefunction contained on the left (right) of the ribbon. Ribbon spectra are calculated for a width of 100 unit cells (600 atomic sites), with a $k$ resolution of $2 \pi / 125$ ($2 \pi / 150$ for the zoomed in regions). Momentum space spectra and Berry curvature evaluated on a $500 \times 500$ grid.}
	\label{fig:Chern_phases}
\end{figure*}
%%%%%%%%%%%%%%%%%%%%%%%%%%%%%%%%%%%%%%%%%%%%%%

\subsection{Chern Number}

The first Chern number is the topological invariant which was first discovered in a physical system and linked to an experiment\,\cite{klitzing-80prl494} (also known as the Thouless--Kohmoto--Nightingale--den Nijs invariant \cite{thouless-82prl405}; it characterizes the (anomalous) quantum Hall effect. It is well-established that for quantum Hall systems the Chern number can be measured in standard transport experiments via transverse (Hall) conductance,
\begin{equation}
\sigma^{xy} = \frac{e^2}{h} C\ .
\end{equation}
A finite Chern number implies broken time-reversal (TR) symmetry (explicitly or spontaneously). For our model \eqref{ham}, TR is explicitly broken by Haldane's term, \ie whenever $t'\not= 0$. A characteristic property of topological phases is the presence of bulk--boundary correspondence. That is, a finite topological invariant causes metallic states, which are topologically protected, to be present at (some of) the edges or surfaces and they must traverse the spectral bulk gap. 

Suppose a band is separated from other bands by an energy gap, the Chern number of that band is defined as
\begin{equation} \label{eqn:Chern_integral}
	c_n = \frac{1}{2 i \pi} \int_{\text{BZ}} F^{(n)}_{xy} (\vec{k}) \cdot d \vec{k}
\end{equation}
where the integral is taken over the first Brillouin zone. $F^{(n)}_{xy}$ is the Berry curvature of the $n$th band,
\begin{equation}
	F^{(n)}_{xy} (\vec{k}) = \frac{\partial A^{(n)}_y (\vec{k})}{\partial k_x} - \frac{\partial A^{(n)}_x (\vec{k})}{\partial k_y}\ .
\end{equation}
The Berry connection $\vec{A}^{(n)} (\vec{k})$ of the $n$th band is defined as
\begin{equation}
	A^{(n)}_i (\vec{k}) = \langle n(\vec{k}) | \frac{\partial}{\partial k_i} | n(\vec{k}) \rangle 
\end{equation}
where $| n (\vec{k}) \rangle $ is the normalised Bloch state of the $n$th band,
\begin{equation}
	h (\vec{k}) | n (\vec{k}) \rangle = E_n | n (\vec{k}) \rangle\ .
\end{equation}
Note that the normalised wavefunctions, and therefore the Berry connection is gauge dependent, however the Chern number is not. 

The Chern number at a particular filling is then the sum of the Chern numbers of the filled bands, 
\begin{equation}
C=\sum_{n\,{\rm occ.}} c_n\ .
\end{equation}
For instance, for \eqref{ham} the Chern number at half filling is $C = \sum_{n = 1}^3 c_n$.

The integral \eqref{eqn:Chern_integral} can be numerically evaluated using the method established in Ref.\,\onlinecite{fukui2005}. We discretize the Brillouin zone into an $N \times N$ grid, defined by reciprocal lattice vectors: $\vec{\mu}_1 = \vec{b}_1/N, \vec{\mu}_2 = \vec{b}_2/N$. At each point in the discretized Brillouin zone $\vec{k}_\ell$, the lattice field strength $F_{12}$ is defined as
\begin{equation}
	F_{12} (\vec{k}_\ell) = \ln ( U_1 (\vec{k}_\ell) U_2 (\vec{k}_\ell + \vec{\mu}_1)  U_1 (\vec{k}_\ell + \vec{\mu}_2)^{-1}  U_2 (\vec{k}_\ell)^{-1} )
\end{equation}
$F_{12} (\vec{k}_\ell) $ is purely imaginary, and Im($F_{12} (\vec{k}_\ell) $) is defined on the principal branch of the logarithm. $U_i (\vec{k}_\ell)$ is the link variable,
\begin{equation}
	U_i (\vec{k}_\ell) = \langle n (\vec{k}_\ell) | n ( \vec{k}_\ell + \vec{\mu}_i) \rangle
\end{equation}
where $| n (\vec{k}_\ell) \rangle$ is the $n$th normalised energy eigenvector at $\vec{k}_\ell$. The Chern number of the $n$th band is then
\begin{equation}
	c_n = \frac{1}{2 \pi i} \sum_{\ell} F_{12} (\vec{k}_\ell)
\end{equation}
where we sum over the entire discretized Brillouin zone.

By virtue of bulk--boundary correspondence, $C$ can also be read off from ribbon spectra such as those in Fig.\,\ref{fig:Chern_phases} (a,c,e). The ribbon spectra are calculated by Fourier transforming the lattice in one direction to momentum space, while keeping the other real space coordinate. The result is a $6N$-dimensional basis system, with one quantum number $k$. For Chern phases with finite Chern number $C$, we expect the difference of right- and left-moving edge modes {\it per edge} to be $C$. For all Chern phases considered in this paper, we have checked that the edge mode counting in ribbon geometry matches the calculated Chern numbers. We note that this procedure applies to any other filling.

%%%%%%%%%%%%%%%%%%%%%%%%%%%%%%%%%%%%%%%%%%%%%%%%%%%%%%%%%%%%%%%%%%%%%%%%
\begin{table}[t!]
	\begin{ruledtabular}
		\begin{tabular}{ccccccccc}
			IV & V & VI & VII & VIII & IX & X & XI & XII \\
			\hline
			1 & 4 & -2 & 2 & -4 & 5 & -1 & 3 & -2 \\
		\end{tabular}
	\end{ruledtabular}
	\caption{Chern Numbers of phases in Fig.\,\ref{fig:phase_boundary_diagram}. Phases I-III possess a zero Chern number.}
	\label{table:Chern_no}
\end{table} 
%%%%%%%%%%%%%%%%%%%%%%%%%%%%%%%%%%%%%%%%%%%%%%%%%%%%%%%%%%%%%%%%%%%%%%%%%%

Phases I-III have a zero Chern number; all other nine phases have a non-zero Chern number, ranging in value from -4 to 5, see Table\,\ref{table:Chern_no}.

\subsection{Discussion of Phases}

We have selected three phases which exemplify the behaviour of the nine Chern phases. Ribbon spectra, spectra in momentum space and Berry curvature for these three phases are the subject of Fig.\,\ref{fig:Chern_phases}; the corresponding plots for the other six phases are delegated to the Appendix, see Figs.\,\ref{fig:appendix_phases_1} and \ref{fig:appendix_phases_2}.

\subsubsection{Phase IV: Haldane phase}

Phase IV is the largest phase in our phase diagram and contains at $\alpha=1$ the original Haldane phase\,\cite{haldane1988}, albeit with a three-times increased unit cell. The Chern number of Haldane's original model is $C=1$, and by adiabatic continuity $C$ cannot change within the entire phase. Here we discuss the phase mainly for benchmarking purposes. A representative ribbon spectrum shown in Fig.\,\ref{fig:Chern_phases}\,(a) clearly highlights one right-moving (left-moving) mode on the left (right) edge at half filling. The sum over the Chern numbers of the three occupied bands [Fig.\,\ref{fig:Chern_phases}\,(b)] yields $C=-1+0+2=1$ as expected. The emerging Chern phases at lower fillings are discussed in Sec.\,\ref{sec:lower-fillings}.

As mentioned before, the phase boundary is determined by gap-closings at the $\Gamma$, $K$, $K'$ or $M$ points in the Brillouin zone. There are a few special points which require our attention: (i) $\alpha=1$, $\lambda=0$ corresponding to undistorted graphene. (ii) $\alpha=0$ and $\lambda = 1/\sqrt{3}$ represents the only phase boundary point which is in the decoupled hexamer limit [the evolution of energies at $\alpha=0$ as a function of $\lambda$ are shown in Fig.\,\ref{fig:slice_spectra_combo}\,(a)]. (iii) $\alpha=1$, $\lambda\to\infty$ corresponds to $t'=1$ and $t=0$, \ie, the honeycomb lattice decouples into its two triangular sublattices. Note that these triangular lattices are subject to imaginary hoppings and thus different from standard triangular lattice bands; in fact, at this point the bandstructure is that of a semi-metal on the triangular lattice\,\cite{rachel-15prl167201}.

\subsubsection{Phase V}

Phase V requires large values of $\alpha$ and significant $\lambda \sim 1$. Energy gaps at half filling turn out to be quite small in the entire phase, see Fig.\,\ref{fig:Chern_phases} (c,d) and in particular the zoom-in in panel (c).
This phase has a Chern number $C=1+0+3=4$. 
Despite the small gap size, the four chiral edge modes can be well observed [Fig.\,\ref{fig:Chern_phases} (c)].

\subsubsection{Phase VII}

\begin{figure}[h!]
\centering
\includegraphics[width=0.95\linewidth]{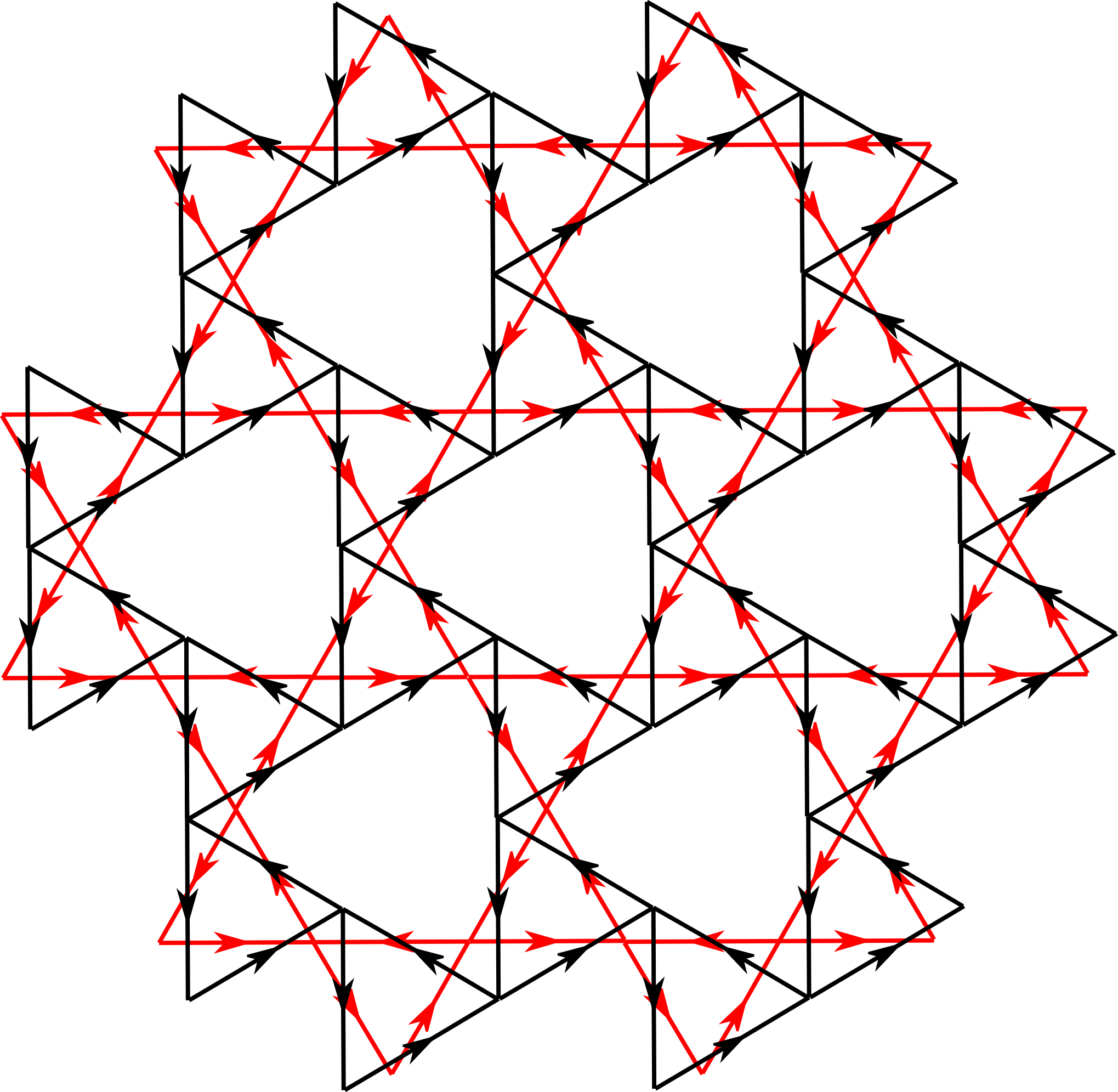}
\caption{Sublattice of the honeycomb lattice for $\alpha\to\infty$ and $\lambda\to\infty$ (black). In red, an equivalent kagome lattice is shown. By slightly shifting the sites and bonds around, one can transform one lattice into the other. Arrows indicate the phase convention of the imaginary hopping term with amplitude $t'$.}
\label{fig:kagome}
\end{figure}

Phase VII is a rather large phase in Fig.\,\ref{fig:phase_boundary_diagram}, realized through parameters $\alpha>2$ and $\lambda>1$. The phase has two flat, topologically trivial bands closer to zero energy; the Chern number is found as $C=1+1+0=2$, in agreement with two chiral edge modes [Fig.\,\ref{fig:Chern_phases} (e,f)]. Moreover, the phase contains the special point $\alpha\to\infty$, $\lambda\to\infty$, albeit gapless due to perfectly flat bands at zero energy. As before, $\lambda\to\infty$ corresponds to decoupled triangular-sublattices of the original honeycomb lattice. Moreover, sending $\alpha\to\infty$ isolates and decouples further sites from the triangular lattices. The remaining graph is adiabatically equivalent 
to a kagome lattice with additional isolated lattice sites. The equivalence to the kagome lattice is illustrated in Fig.\,\ref{fig:kagome}.

%%%%%%%%%%%%%%%%%%%%%%%%%%%%%%%%%%%%%%%%%%%%%%%%%%%%%%%%%%%%%%%
%%%%%%%%%%%%%%%%%%%%%%%%%%%%%%%%%%%%%%%%%%%%%%%%%%%%%%%%%%%%%%%

The kagome lattice tight binding model with real nearest-neighbor hopping appears to be the one of the honeycomb lattice with an additional perfectly flat band touching the other bands at the top or bottom, depending on the sign of the nearest-neighbor hopping amplitude. Here the nearest-neighbor term is, however, purely imaginary (Haldane's term). The bands appear like gapped graphene, but with a flat band in between at zero energy. 

The spectrum is identical to that of a kagome tight binding model with staggered fluxes applied\,\cite{ohgushi2000,green-10prb075104}, causing topologically non-trivial behavior. The staggered fluxes lead to a Chern number $C=1$ in the gaps above and below the flat zero-energy band.
Perturbing slightly away from this special point, leads to a weakly coupling of the two kagome lattices (for finite $\lambda < \infty$) or a weak coupling of the isolated atoms in the center of the kagome-hexagons (for finite $\alpha < \infty$), or both. 
Coupling both triangular sublattices, leads then to a total Chern number $C=2$ as found before; also the flat zero-energy bands are hybridized to finite energy, rendering the phase gapped at half filling.
The weak coupling of the two sublattices likely forms a symmetric and an anti-symmetric superposition of the original two sublattice degrees of freedom. The symmetric edge state then overlaps slightly with an anti-symmetric bulk band, and vice versa. These edge state therefore do not hybridize with the overlapping bulk bands [see Fig.\,\ref{fig:Chern_phases}\,(e)].

%%%%%%%%%%%%%%%%%%%%%%%%%%%%%%%%%%%%%%%%%%%%%%%%%%%%%%%%%%%%%%%%%%%%%%%%%%%

%%%%%%%%%%%%%%%%%%%%%%%%%%%%%%%%%%%%%%%%%%%%%%%%%%%%%%%%%%%%%%%
%
%                                            1 / 2 F I L L I N G    :   H O T I
%
%%%%%%%%%%%%%%%%%%%%%%%%%%%%%%%%%%%%%%%%%%%%%%%%%%%%%%%%%%%%%%%
\section{Higher Order Topological Phases at half filling}
\label{sec:hoti-phases}

%%%%%%%%%%%%%%%%%%%%%%%%%%%%%%%%%%%%%%%%%%%%%%
\begin{figure*}[phtb]
	\centering
	\includegraphics[width=\textwidth]{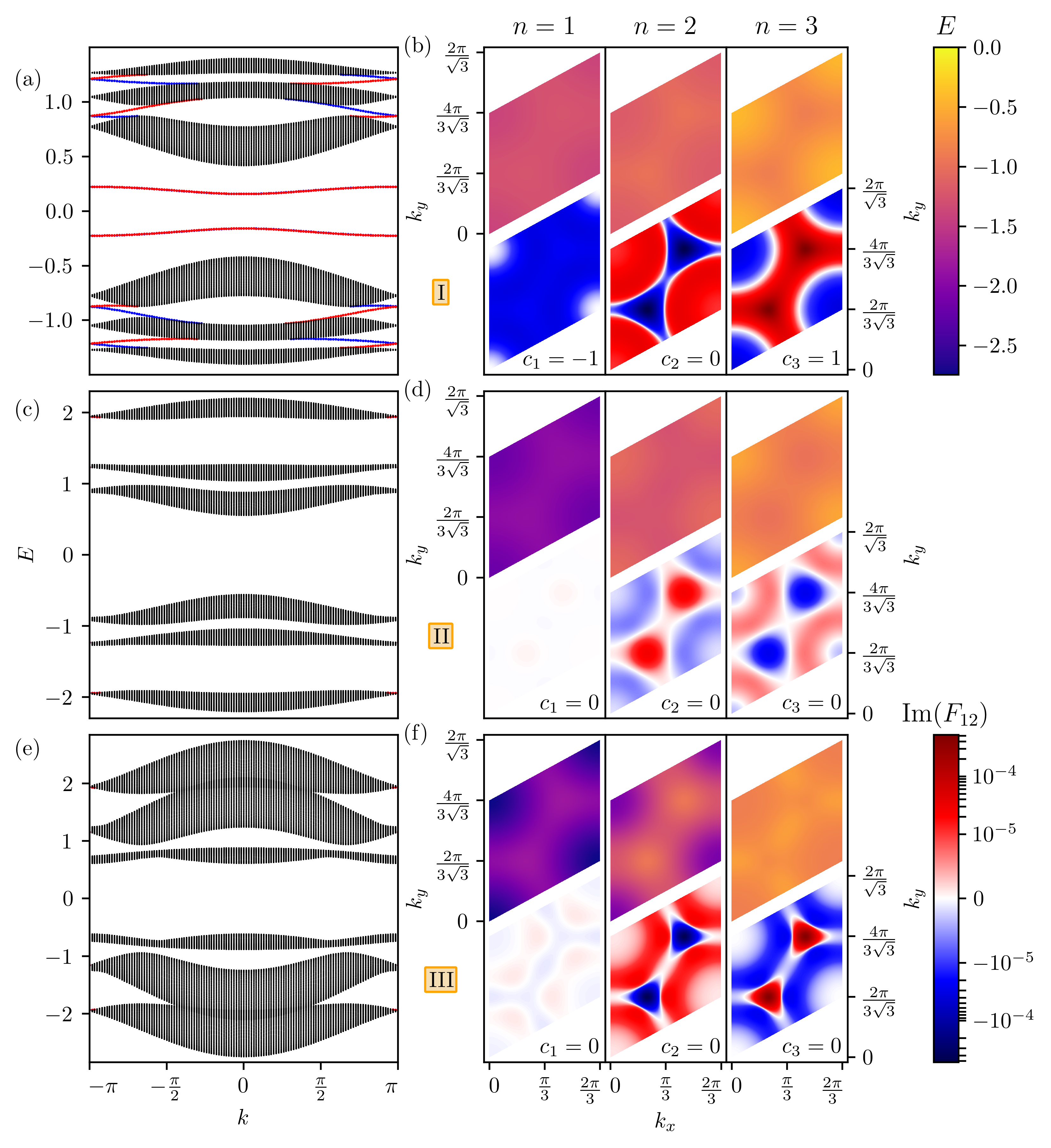}
	\caption{Ribbon spectra (a, c, e), spectra in momentum space (top) and Berry curvature (bottom) over the Brillouin zone (b, d, f). Chern numbers for each of the three cases shown is $C=0$ at half filling. First row (a, b) is phase I, parameters used $\alpha = \frac{1}{0.2}, \lambda = 0.1$; second row (c, d) is phase II, parameters used $\alpha = 0.2, \lambda = 0.1$; third row (e, f) is phase III, parameters used $\alpha = 0.2, \lambda = \frac{1}{0.4}$. Displayed below Berry curvature plots are the Chern numbers $c_n$ of each individual band. For the ribbon spectra, blue (red) dots are states with $\geq$75\% of the wavefunction contained on the left (right) of the ribbon. Ribbon spectra are calculated for a width of 100 unit cells (600 atomic sites), with a $k$ resolution of $2 \pi / 125$. Momentum space spectra and Berry curvature evaluated on a $500 \times 500$ grid.}
	\label{fig:hoti_phases}
\end{figure*}
%%%%%%%%%%%%%%%%%%%%%%%%%%%%%%%%%%%%%%%%%%%%%%

Higher $n$th-order topology for a $d$-dimensional system is evident as a bulk-boundary correspondence with $d-n$ edge or surface states, respectively. Just like a 2D first-order topological insulator possesses one-dimensional edge states traversing the bulk gap, a 2D second-order topological insulator possesses zero-dimensional {\it corner} modes localized within the gap\,\cite{benalcazar2017}. The Benalcazar--Bernevig--Hughes (BBH) model for the $\pi$-flux square lattice\,\cite{benalcazar2017} translates to the honeycomb lattice with Kekul\'e or anti-Kekul\'e distortion without the need to apply any flux\,\cite{mizoguchi2019}. The honeycomb HOTI phases (for $\lambda=0$) were previously characterized by virtue of $\mathbb{Z}_Q$ Berry phases which were introduced in Refs.\,\cite{hatsugai2006, hatsugai2011, mizoguchi2019, araki2020}. In the following, we extend this pioneering work and investigate the entire $\alpha$--$\lambda$ phase diagram and search for HOTI phases.

Irreducible cluster representations, \ie configurations where the lattice is decoupled into monomers, dimers, trimers etc., are key to understand HOTI phases. For instance, the previously discussed anti-Kekul\'e phase II contains the decoupled hexamer point $\alpha=0=\lambda$. As long as the bulk gap remains finite, the HOTI phase will persist and the $\mathbb{Z}_Q$ invariant to stay constant. Similarly, $\alpha\to\infty$ and $\lambda=0$ is the decoupled dimer point of the Kekul\'e phase I\,\cite{mizoguchi2019}.

The $\mathbb{Z}_Q$ Berry phase\,\cite{hatsugai2011, mizoguchi2019, araki2020} can take $Q$ discrete values. Due to the (anti-) Kekul\'e distortion causing dimerization or hexamerization, respectively, we expect only $Q=2$ and $Q=6$ to be of relevance; thus we compute the $\mathbb{Z}_2$ and $\mathbb{Z}_6$ invariants for the entire phase space.

In agreement with the irreducible cluster representations mentioned before, we find phase I to be a $\mathbb{Z}_2$ HOTI phase with $\mathbb{Z}_2=2\pi/2\times 1 = \pi$ and phase II to be $\mathbb{Z}_6$ HOTI phase with $\mathbb{Z}_6=2\pi/6\times 3 = \pi$. In addition, we identify phase III also as a $\mathbb{Z}_6$ HOTI  with $\mathbb{Z}_6=2\pi/6\times 3 = \pi$. Ribbon spectra, spectra in momentum space, and Berry curvature for these three phases are the subject of Fig.\,\ref{fig:hoti_phases}.
Naively one might assume first- and second-order topologies are exclusive, \ie either a phase is one or the other. Thus one would also assume the Chern phases should possess $\mathbb{Z}_Q=0$. In contrast to such reasoning, we find finite $\mathbb{Z}_2$ and $\mathbb{Z}_6$ Berry numbers within some of the Chern phases, albeit not adiabatically connected to irreducible cluster representations. Even more surprising, we find discrete transitions of the Berry phases within Chern phases (\ie without closing of the bulk gap), as discussed below.

\subsection{$\mathbb{Z}_2$ Berry Phase}
\label{sec:Z2_Berry_phase}

%%%%%%%%%%%%%%%%%%%%%%%%%%%%%%%%%%%%%%%%%%%%%%
\begin{figure}[ht]
	\begin{center}
		\includegraphics{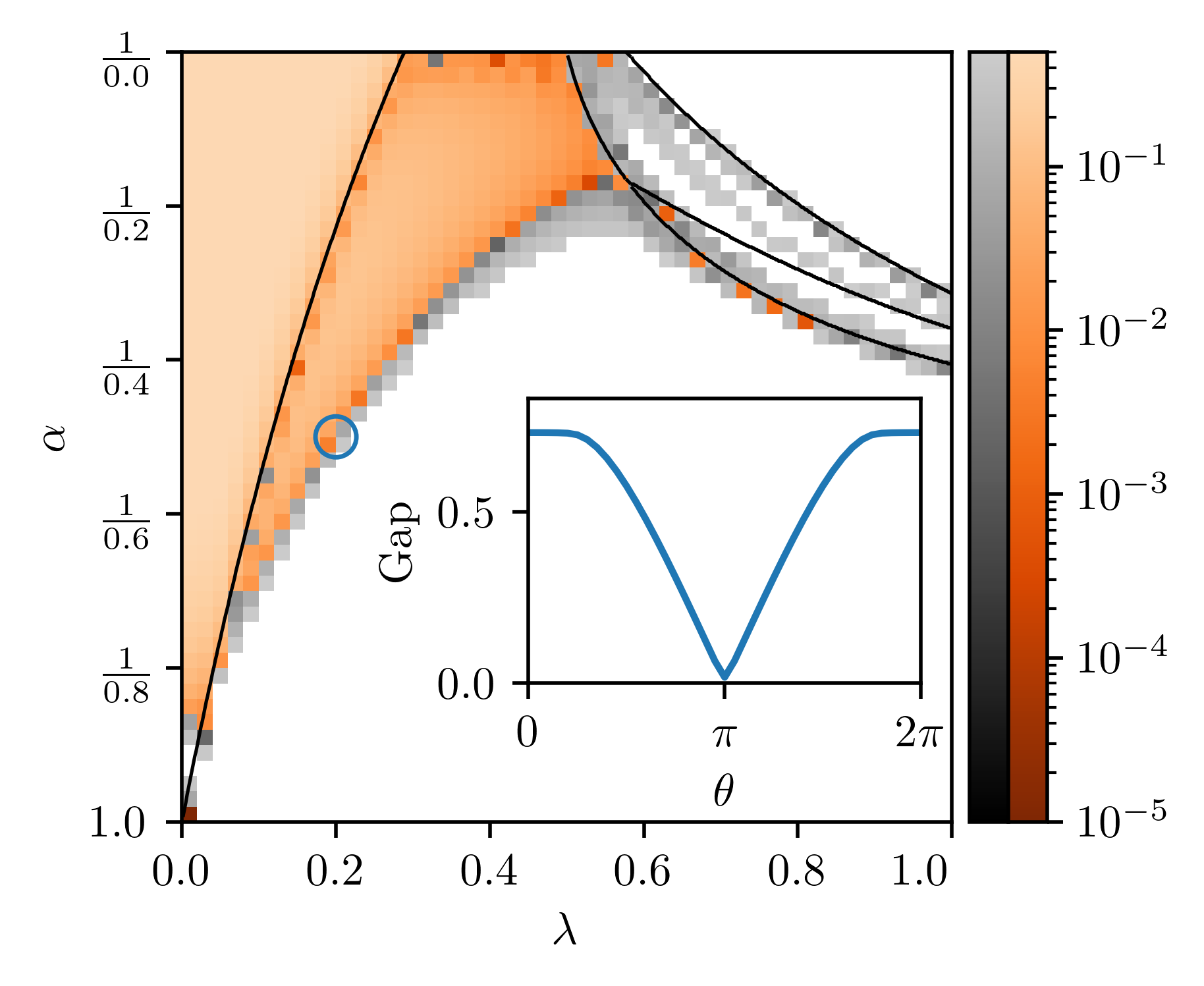}
		\caption{
			The $\mathbb{Z}_2$ Berry phase diagram. Orange is a Berry phase of $\pi$, and white/black is a Berry phase of $0$. The colour scales correspond to the smallest energy gap of $H(\theta)$ along the Berry phase integral. Calculated for $N=18$ (\ie $6\times 18^2$ lattice sites). The $\mathbb{Z}_2$ Berry phase is zero in all other phases. The half filling phase boundaries from Fig.\,\ref{fig:phase_boundary_diagram} have been overlaid as black lines. Inset is the half-filling energy ``twist'' gap parameters for $\alpha = \frac{1}{0.5}, \lambda = 0.2$ (blue circle), which highlights that the gap is closing for twist angle $\theta = \pi$; see main text for discussion.
		}
		\label{fig:half_fill_z2}
	\end{center}
\end{figure}
%%%%%%%%%%%%%%%%%%%%%%%%%%%%%%%%%%%%%%%%%%%%%%

%%%%%%%%%%%%%%%%%%%%%%%%%%%%%%%%%%%%%%%%%%%%%%
\begin{figure}[ht]
	\begin{center}
		\includegraphics[width=0.48\textwidth]{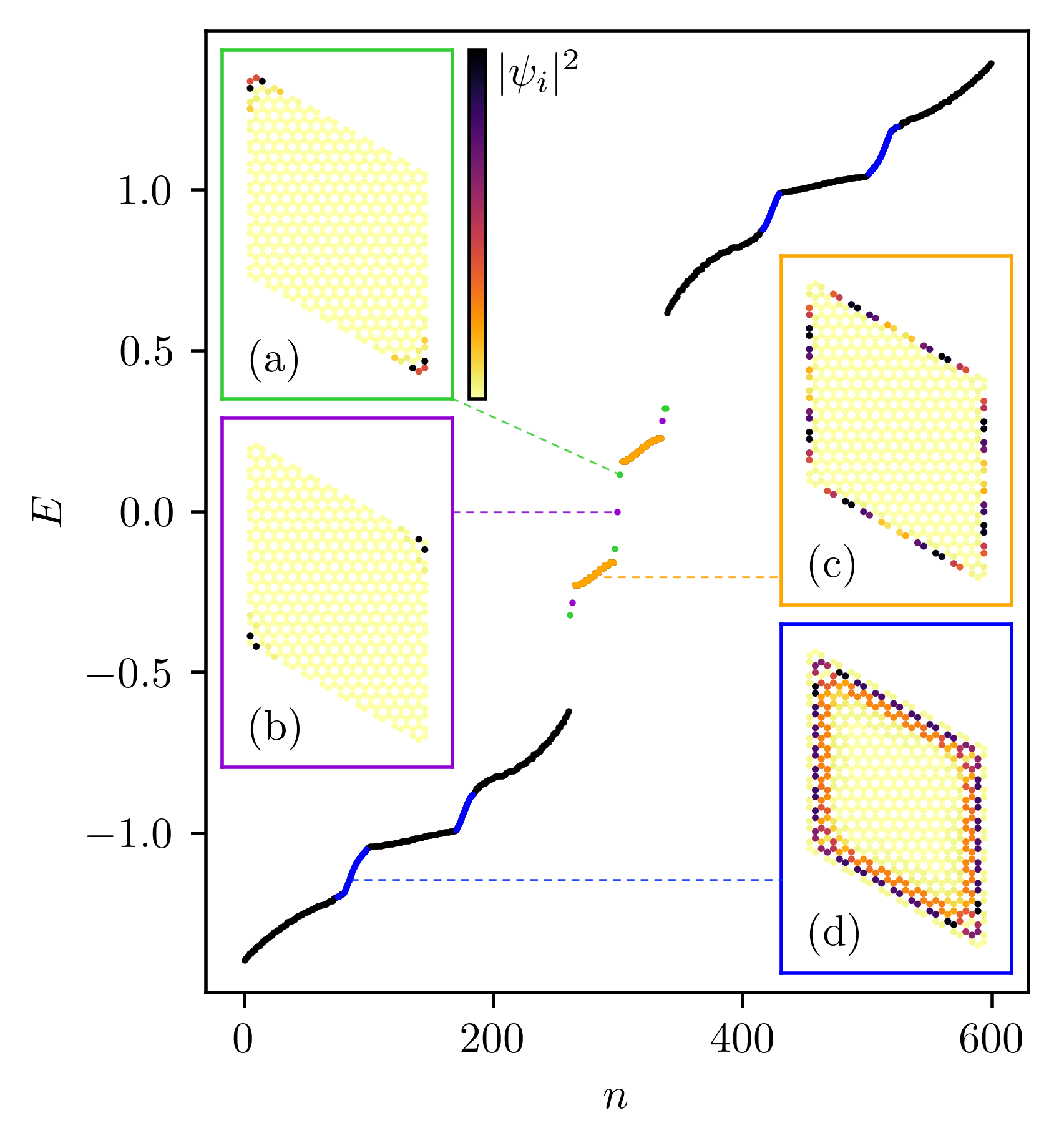}
		\caption{Example of energy spectrum in phase I ($\alpha = \frac{1}{0.2}, \lambda = 0.05$, on a $10 \times 10 \times 6 - 2 = 598$ site lattice). Energies are plotted in increasing order $n$. Wavefunctions of interest are shown in the inset boxes corresponding to their colour. (a) $|\psi|^2$ of tetramer structure corner state at $E=0.1162$. (b) $|\psi|^2$ of trimer structure corner state at $E=0$, \ie one of the two zero-energy corner states at half filling. Note that these two wavefunctions look identical. (c) $|\psi|^2$ of dimer edge state at $E=-0.2262$. (d) $|\psi|^2$ of chiral edge state in the sixth filling gap at $E=-1.1578$. Chiral edge states exist in the third and sixth filling gaps, which both have a Chern number of $-1$. Yellow (black) corresponds to zero (high) wavefunction density in (a)-(d).}
		\label{fig:real_space_geometry_z2}
	\end{center}
\end{figure}
%%%%%%%%%%%%%%%%%%%%%%%%%%%%%%%%%%%%%%%%%%%%%%

Consider the dimer cluster limit $\alpha \rightarrow \infty, \lambda = 0$ adiabatically connected to phase I, where the only non-zero bonds are the nearest neighbour bonds, \ie dimers, between the hexamer. Dimers have a $\mathbb{Z}_2$ sublattice symmetry, thus we consider the $\mathbb{Z}_2$ Berry phase.

We choose one of the disconnected dimer bonds, and w.l.o.g. label its lattice sites 1 and 2. We may separate the Hamiltonian as
\begin{align}
	H &= H_\text{dimer} + (H - H_\text{dimer}) \\
	&= c_1^\dagger c_2 + c_2^\dagger c_1 + (H - H_\text{dimer})\ .
\end{align}
Next we introduce a ``twist’’ on this dimer bond by transforming the operators corresponding to one of the dimer sites,
\begin{equation} \label{eqn:twist_factor_z2}
	c_1 \mapsto e^{i \theta} c_1
\end{equation}
while leaving the other one unchanged. Thus the Hamiltonian becomes a function of $\theta$:
\begin{align}
	H(\theta) &= H_\text{dimer} (\theta) + (H - H_\text{dimer}) \\
	&= e^{-i \theta} c_1^\dagger c_2 + e^{i \theta} c_2^\dagger c_1 + (H - H_\text{dimer})
\end{align} 
Since the energy eigenstates $ | n (\theta) \rangle $ are often degenerate, we need to compute the trace over the matrix $ \langle n (\theta) | \frac{\partial}{\partial \theta}  | m (\theta) \rangle$ instead of the standard Berry potential, thus sometimes also referred to as Non-Abelian Berry phase\,\cite{hatsugai2004, hirano2008}. The non-Abelian Berry phase accumulated as we `twist' this dimer through $2 \pi$ is
\begin{equation} \label{eqn:nonab_berry_z2}
	\gamma_2 = \int_0^{2 \pi} \text{Tr } A (\theta) d \theta
\end{equation}
where $A (\theta)$ is the non-Abelian Berry connection\,\cite{hirano2008}:
\begin{equation} \label{eqn:nonab_berry_conn}
	A (\theta) = - i \Psi^\dagger (\theta) \frac{\partial}{\partial \theta} \Psi  (\theta)\ .
\end{equation}
$ \Psi(\theta)$ is a multiplet, defined as
\begin{equation}
	\Psi(\theta) = [ | 1 (\theta) \rangle , | 2 (\theta) \rangle , ... , | M (\theta) \rangle ]\ ,
\end{equation}
\ie a matrix whose columns are the lowest $M$ normalised energy eigenvectors $| n (\theta) \rangle$,
\begin{equation}
	H (\theta) | n (\theta) \rangle = E_n (\theta) | n (\theta) \rangle
\end{equation}
and $M$ is the number of filled states. For instance, $M = N/2$ for half filling, where $N$ is the number of lattice sites.

If we consider the integration path in Eq.\,\eqref{eqn:nonab_berry_z2} as half of a closed loop from $\theta = 0$ to $\theta = 2 \pi$ and back again, then the condition that the gauge must be regular (single-valued) demands that the Berry loop over the whole closed loop must be equal to an integer multiple of $2 \pi$. However, the integral for $\theta$ from $0$ to $2 \pi$ is identical to the integral from $2 \pi$ back to $0$. The latter integral is the equivalent of the former integral if we were to apply the gauge transformation \eqref{eqn:twist_factor_z2} to the other atom in the dimer $c_2$ rather than $c_1$. The $\mathbb{Z}_2$ symmetry of the dimer forces the two integrals to be identical. Therefore, the Berry phase defined in Eq.\,\eqref{eqn:nonab_berry_z2} must be equal to 0 or $\pi$ modulo $2 \pi$, \ie the $\mathbb{Z}_2$ twist Berry phase is quantized\,\cite{hatsugai2011}. 

Numerically, we can calculate the non-Abelian Berry phase by breaking up the integration path into $N+1$ discrete values $\theta_i$, $\{ \theta_i = 2 \pi i / N \, | \, 0 \leq i \leq N \}$. We can then use a lattice Berry connection so Eqn.\,\eqref{eqn:nonab_berry_conn} at $\theta = \theta_i$ becomes
\begin{equation}
	A_i =	- i \Psi^\dagger (\theta_i)  \Psi (\theta_{i+1})\ .
\end{equation} 
The Berry phase \eqref{eqn:nonab_berry_z2} can then be evaluated as
\begin{equation}
	\gamma_2 =  i \prod_{i=0}^{N} \text{Arg } \det A_i\ .
\end{equation}
Note that this calculation is gauge-invariant as long as we start and end the integration at the same point, which is the case in our calculation\,\cite{king-smith_vanderbilt_1993, hatsugai2004}.

Up to this overall gauge freedom, these Berry phases are well defined and do not change unless the spectral gap, depending on twist angle $\theta$, closes. There are two ways in which the gap may close: (i) the gap closes with no twist and is thus identical to the bulk gap (as it happens for the boundaries of the phases I and II in Fig.\,\ref{fig:phase_boundary_diagram}); (ii) the gap remains open for zero twist but closes for a finite twist angle. In the $\mathbb{Z}_2$ twist Berry phase calculation, the latter case only occurred when $\theta = \pi$, where the $\mathbb{Z}_2$ twist Berry phase transitions between $\pi$ and 0 away from the phase boundaries through the middle of phase IV (see Fig.\,\ref{fig:half_fill_z2}).

The numerical calculation shown in Fig.\,\ref{fig:half_fill_z2} has two phases where the $\mathbb{Z}_2$ twist Berry phase $ \gamma_2 = \pi$. One coincides with phase I, where the Chern number is zero, so the constant and non-zero $\mathbb{Z}_2$ Berry phase characterises the phase. The other phase overlaps some of phase IV, where the Chern number is one. Corner states that sit in the bulk energy gap at $E=0$ can be observed in phase I. Where there is a finite Chern number in the half-filling gap, corner states cannot be discerned from the edge states which are linearly dispersed across the bulk gap. Phases with a finite Chern number and finite $\mathbb{Z}_Q$ Berry phase are discussed further in Sec.\,\ref{sec:discussion}.

The presence of corner states in phase I is dependent on the edge geometry. There are many possible geometries that give corner states, but a requirement is that there is a complete hexamer on a corner that is attached to either one, three, or five other hexamer unit cells\,\cite{mizoguchi2019, liu2019, zangeneh-nejad2019, lee2020, li2020}. Consider the limit where inter-hexamer bonding is much stronger than a small (but finite) intra-hexamer bond. Then, all atoms in the hexamer that are connected to a different hexamer in the lattice are involved in forming a strong inter-hexamer dimer, and are practically disconnected from the other atoms in the hexamer. The atoms that are not dimerized in this way then form a free monomer, trimer or pentamer with a zero energy eigenstate.   An example geometry is considered in Fig.\,\ref{fig:real_space_geometry_z2} (which is reminiscent to that shown in Ref.\,\onlinecite{li2020}) where the 120 degree corner hexamers are connected to three other hexamer unit cells each, so a weakly connected trimer is formed. The inset (b) shows the zero-energy corner states are localised on these corners. Note that isolated trimer (and pentamer) states feature additional energy states different from $E=0$. Where corner hexamers are attached to two other hexamer unit cells, a tetramer is formed, which are the green energy eigenstates in the diagram. These tetramer structures (and dimer structures, like those on the edge of the lattice) do not have zero energy states.

\subsection{$\mathbb{Z}_6$ phase}
\label{sec:Z6_Berry_phase}

%%%%%%%%%%%%%%%%%%%%%%%%%%%%%%%%%%%%%%%%%%%%%%
\begin{figure*}[ht]
	\centering
	\includegraphics[width=0.48\textwidth]{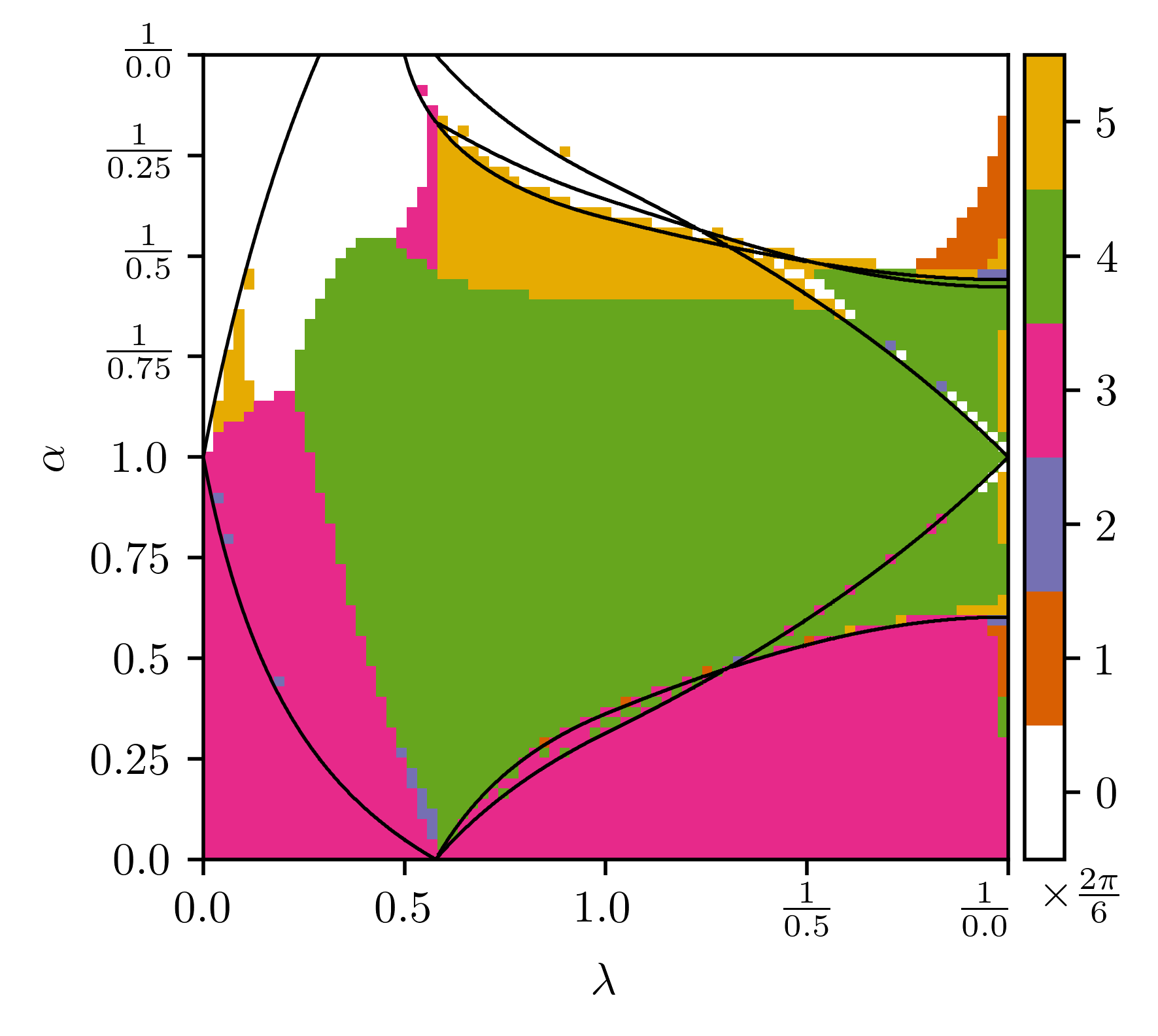}
	\includegraphics[width=0.48\textwidth]{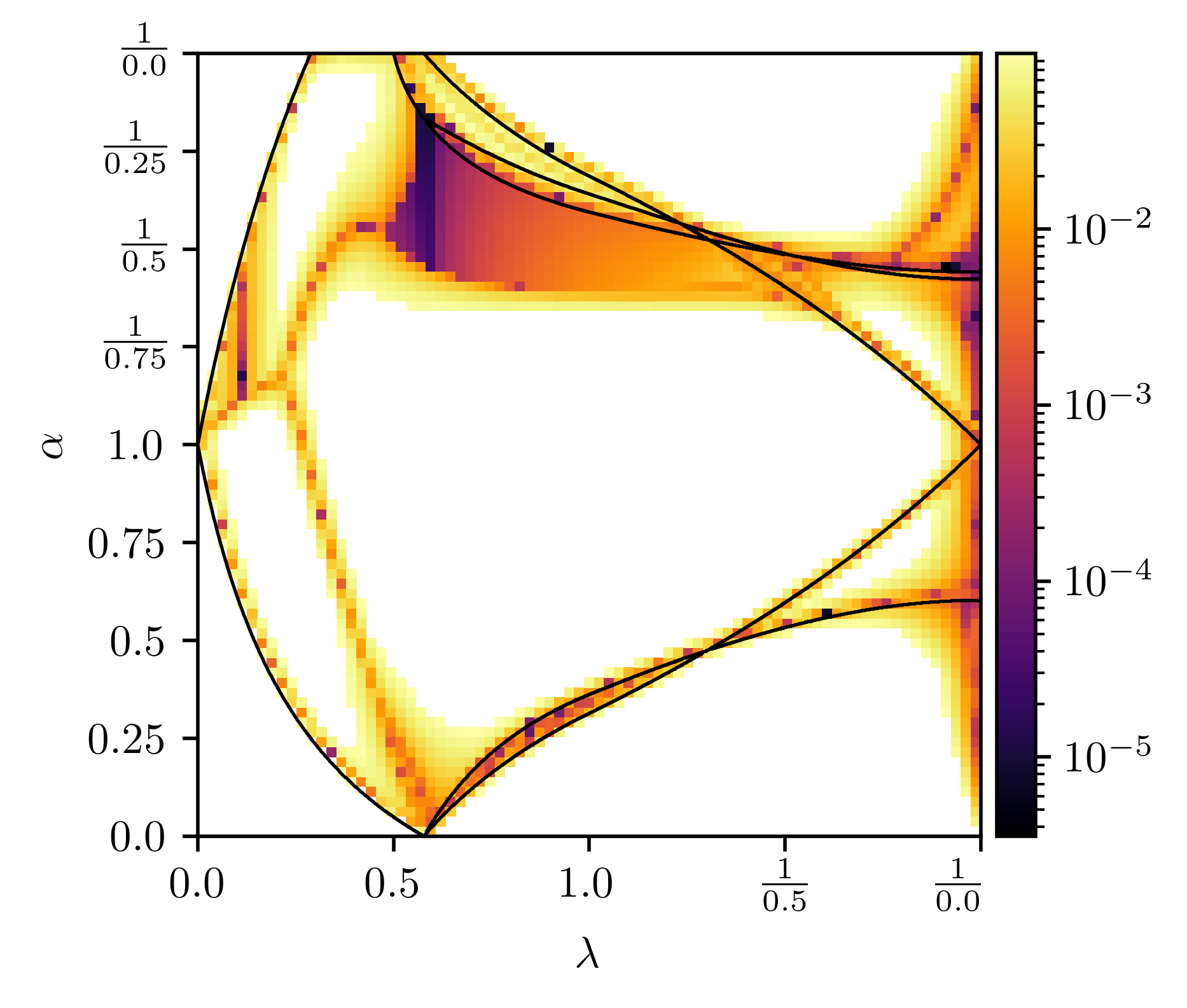}
	\caption{(Left) $\mathbb{Z}_6$ twist Berry phase at half filling. (Right) Minimum energy gap along the twist path. Calculated for $N=18$ real-space lattice (\ie $6\times 18^2$ sites). The half-filling phase boundaries from Fig.\,\ref{fig:phase_boundary_diagram} are overlaid as black lines.}
	\label{fig:half_fill_z6}
\end{figure*}
%%%%%%%%%%%%%%%%%%%%%%%%%%%%%%%%%%%%%%%%%%%%%%
%%%%%%%%%%%%%%%%%%%%%%%%%%%%%%%%%%%%%%%%%%%%%%
\begin{figure*}[ht]
	\begin{center}
		\includegraphics[width=0.48\textwidth]{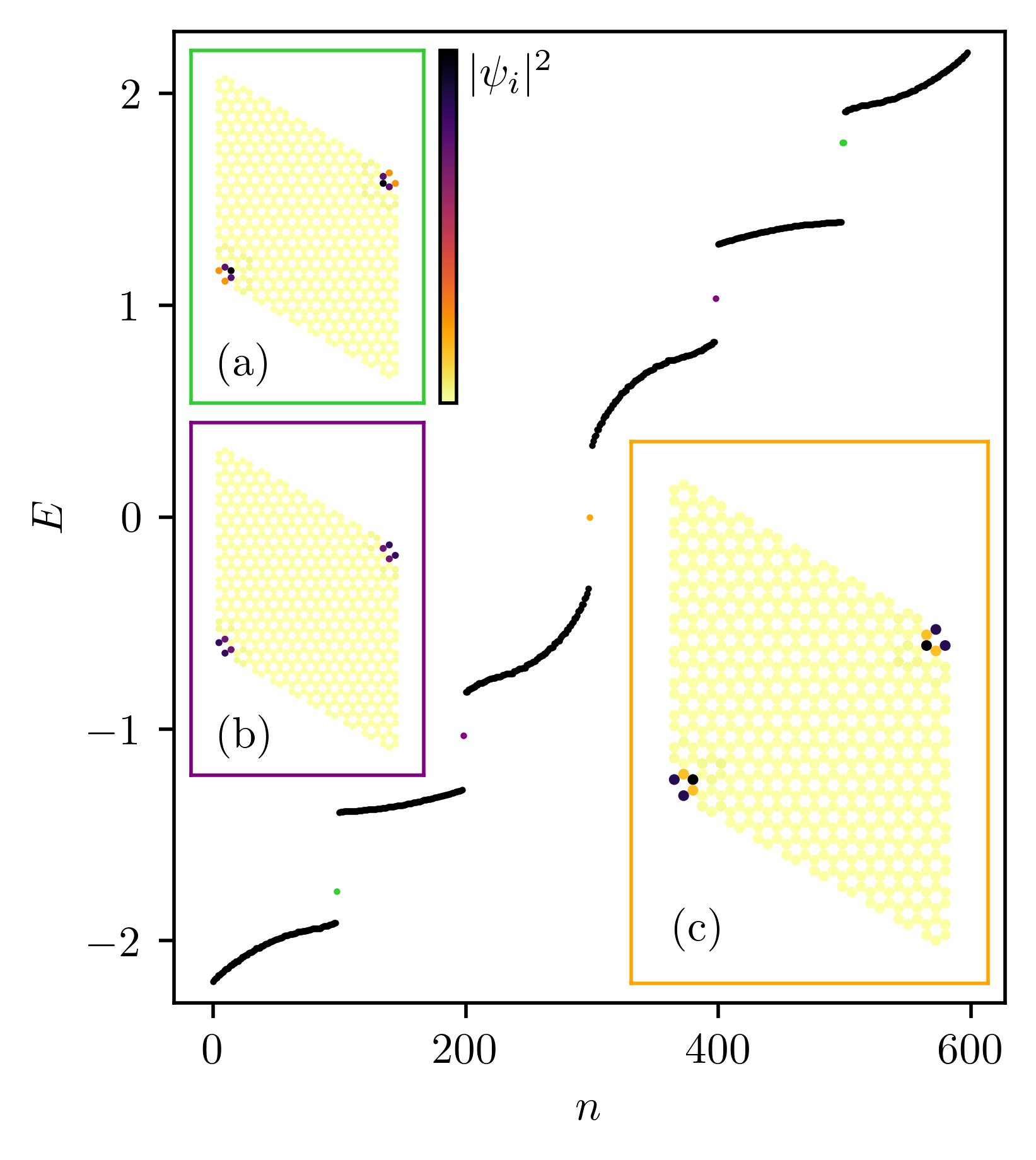}
		\includegraphics[width=0.48\textwidth]{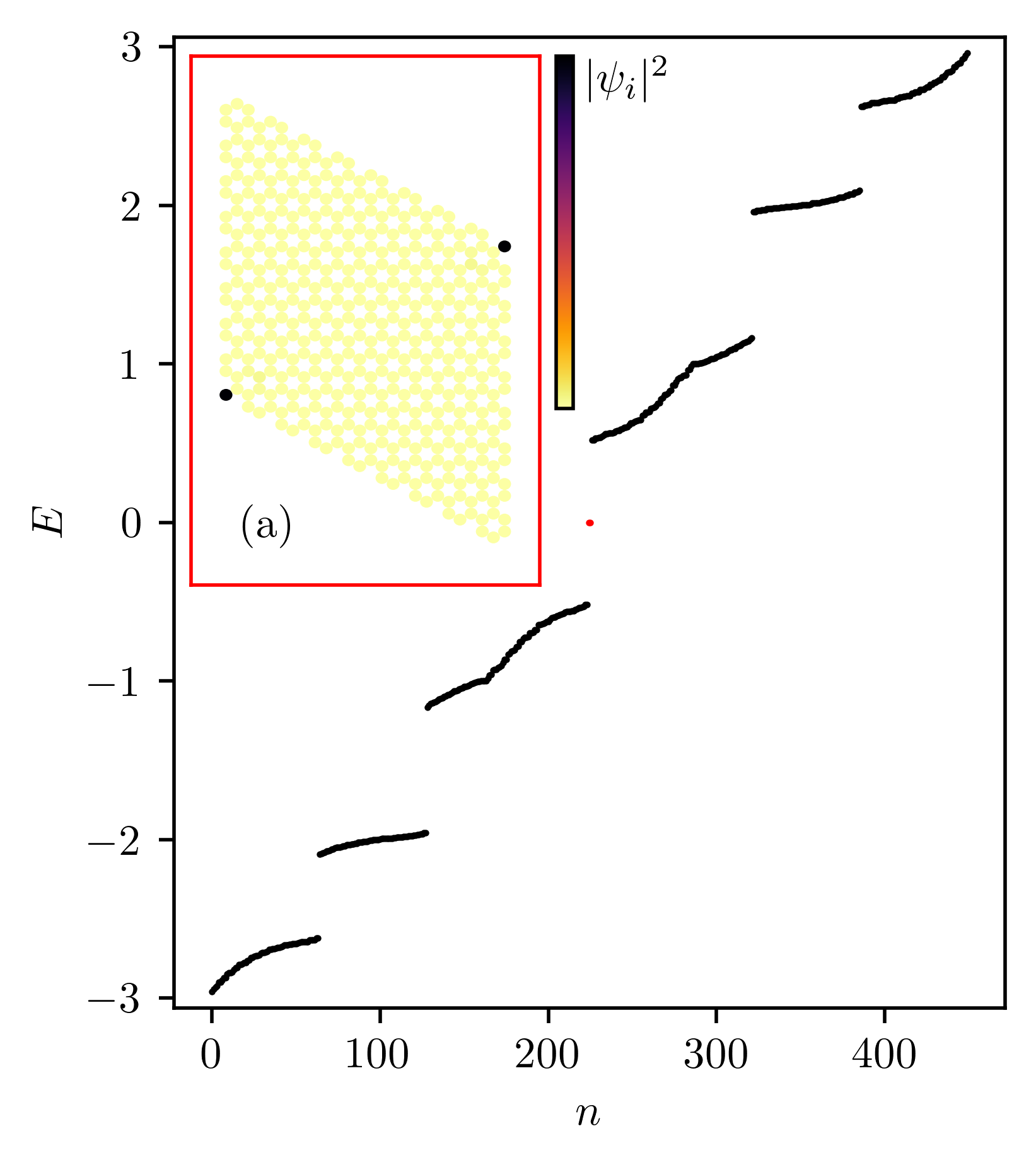}
		\caption{Example of energy spectra. (Left) Phase II ($\alpha = 0.2, \lambda = 0.2$ and, on a $10 \times 10 \times 6 - 2 = 598$ site lattice). The pairwise degenerate states within the different filling gaps are highlighted in various colors and correspond to the corner states shown in insets with the same color. (a) $|\psi|^2$ of an in-gap corner state at 1/6 filling. (b) $|\psi|^2$ of an in-gap corner state at 1/3 filling. (c) $|\psi|^2$ of an in-gap zero-energy corner state at half filling. (Right) Phase III ($\alpha = 0.2, \lambda = 1$ on a 550 site lattice). Energies are plotted in increasing order $n$. The pairwise degenerate states within the half filling gap are highlighted in red and correspond to the $|\psi|^2$ of the corner state shown in inset (a) $|\psi|^2$ of in-gap corner state at 1/2 filling. Yellow (black) corresponds to zero (high) wavefunction density in the insets.}
		\label{fig:real_space_geometry}
	\end{center}
\end{figure*}
%%%%%%%%%%%%%%%%%%%%%%%%%%%%%%%%%%%%%%%%%%%%%%

Now consider the decoupled hexamer cluster limit $\alpha = 0$. Each individual hexamer has a $\mathbb{Z}_6$ symmetry (rotating the labels of the atoms in a hexamer), thus we consider the $\mathbb{Z}_6$ Berry phase. The construction is similar to that of the $\mathbb{Z}_2$ Berry phase, but instead of considering a dimer in the decoupled dimer limit, we consider a hexamer in the decoupled hexamer limit. W.l.o.g. we choose a hexamer, and split the Hamiltonian $H$ into two parts. $H_\text{hexamer}$ contains all bonds between two atoms in the hexamer, and $H-H_\text{hexamer}$ contains all other bonds (bonds between hexamer and any atoms not in the hexamer, and bonds entirely not in the hexamer). 

We then introduce ``twists'' onto the hexamer by transforming the operators in $H_\text{hexamer}$ as
\begin{equation} 
	c_j \mapsto e^{i \phi_j} c_j
\end{equation}
for $j=1,\ldots, 6$ with $\phi_j = \Sigma_{i=1}^j \theta_i$ and $\phi_6 = 0$. The individual $\theta_i$ are the ``twists'' on the nearest neighbour bonds between atoms $i$ and $i+1$ in the hexamer (where $\theta_6$ is the twist on the bond between atom 1 and 6). Note $\theta_6 = - \Sigma_{i=1}^5 \theta_i$, since $\phi_6 = 0$. The transformation also introduces twists on the next nearest neighbour bonds within the hexamer, where the transformation ensures that the net phase accumulated in any closed loop within the hexamer is constant for any amount of ``twisting''.

This transformation renders the Hamiltonian a function of $\Theta = (\theta_1 , ... , \theta_5)$. The non-Abelian Berry phase accumulated as we twist the hexamer over a path $L$ in the 5-dimensional parameter space is,
\begin{equation} \label{eqn:nonab_berry_z6}
	\gamma_6 = \int_L \text{Tr} A (\Theta) \cdot d \Theta
\end{equation}
where $A (\Theta)$ is the non-Abelian Berry connection \eqref{eqn:nonab_berry_conn}.

The path $L$ is any one of the paths $L_i, \, 0 \leq i \leq 5$, where $L_i$ is defined as
\begin{equation}
	L_i : E_i \rightarrow G \rightarrow E_{i+1}
\end{equation}
with $E_0 = E_6 = (0,0,0,0,0), \, E_i = 2 \pi \hat{e}_i$ for $ 1 \leq i \leq 5$, where $\hat{e}_i$ is the unit vector in the $i$th direction, and $G = \frac{2 \pi}{6} (1, 1, 1, 1, 1)$ is the centre of mass of a five dimensional tetrahedron with side lengths of $2\pi$. As implied by the definition of the Berry phase as the integral over any of these paths, these paths are equivalent. This stems from the $\mathbb{Z}_6$ symmetry of rotating the labels of the hexamer, which is equivalent to rotating through the $\theta_i$. 

Since the Berry phase accumulated by taking all 6 paths (which is a closed loop) must be equal to an integer multiple of $2 \pi$, and the 6 paths individually are equivalent to each other, then the $\mathbb{Z}_6$ Berry phase Eq.\,\eqref{eqn:nonab_berry_z6} must be equal to an integer multiple of $2 \pi / 6 = \pi / 3$. The $\mathbb{Z}_6$ twist Berry phase is unchanging unless the gap closes, which, similar to the $\mathbb{Z}_2$ calculation, happens with either no twist, or when $\Theta = G$, the centre of mass point. 

The numerical calculation of the $\mathbb{Z}_6$ Berry phase is analogous to the calculation of the $\mathbb{Z}_2$ Berry phase, see Fig.\,\ref{fig:half_fill_z6} (left). We have observed a finite-size dependence of the results. For instance, the $\gamma_6 = \pi$ phase at $\alpha \approx \frac{1}{0.4}, \lambda \approx 0.6$ on the left in Fig.\,\ref{fig:half_fill_z6}, which is calculated on an $18 \times 18$ hexamer lattice, is not present when the $\mathbb{Z}_6$ Berry phase is calculated on a $17 \times 17$ hexamer lattice. The energy gap diagram on the right in Fig.\,\ref{fig:half_fill_z6} shows that the energy gap is small across this entire phase, which may explain the discrepancy. Similarly, the energy gap is small for the $\gamma_6 = 5 \pi / 3$ phase next to it, and the size of this phase decreases and increases depending if the size of the lattice is $n \times n$ hexamer unit cells with $n$ odd or even. 

There are three phases with $\mathbb{Z}_6=4\pi/3$, covering a large part of phase IV (Haldane phase), and phases X and XI. These phases all have finite Chern numbers ($C=1, -1, 3$ respectively). Analogous to the $\mathbb{Z}_2$ Berry phase, where there was no change in the Berry phase from phase I to the Haldane phase, in the $\mathbb{Z}_6$ Berry phase calculation a Berry phase of $\gamma_6 = \pi$ extends from phase II to the middle of the Haldane phase. This makes a phase with $\gamma_6 = \pi$ and a Chern number with $C=1$. For further discussion of the interplay of $\mathbb{Z}_Q$ Berry phases and Chern numbers, see Sec.\,\ref{sec:discussion}.

\subsubsection{Real space plots/Corner States}

Both phases II and III have a $\mathbb{Z}_6$ twist Berry phase $\gamma_6 = \pi$, and a Chern number of zero.  As $\alpha = 0$ is a decoupled hexamer cluster limit for all $\lambda$, both phases II and III are adiabatically connected to the decoupled cluster limit. (Note that for the analogous situation $\alpha \rightarrow \infty$ only $\lambda = 0$ is a decoupled dimer cluster limit, as second nearest neighbour hopping with finite $\lambda$ couples all the dimers). Like phase I, the specific geometry of the lattice edges and corners determines where zero energy corner states may appear. 

In a similar fashion to the Su-Schrieffer-Heeger model\,\cite{su1979}, where a zero energy edge mode exists when there are lone atoms at the end of a chain of dimers, an incomplete strongly bound hexamer unit cell at the boundary are needed for corner states to exist. 
The simplest case would be a single atom  leading to a $E=0$ state \cite{mizoguchi2019}. Another way to obtain corner states is to remove one atom from a complete hexamer at corners to create two open pentamers (see insets in Fig\,\ref{fig:real_space_geometry} for geometry). 
The pentamer possesses five energy states which turn out to be in-gap states in the five gaps corresponding to 1/6, 1/3, 1/2, 2/3, 5/6 fillings, and are highlighted in color in Fig.\,\ref{fig:real_space_geometry}. The other geometry is a trimer at the corners, also leading to three in-gap states. In contrast, dimers and tetramers at the corner do not create corner states at zero energy, since in the decoupled cluster limit dimers or tetramers do not have a zero energy state. Monomer, trimer or pentamers corners persist across phases II and III until the gap closes, see spectra in Fig.\,\ref{fig:slice_spectra_combo} of the Appendix.

%%%%%%%%%%%%%%%%%%%%%%%%%%%%%%%%%%%%%%%%%%%%%%%%%%%%%
%
%                                                    L O W E R    F I L L I N G S
%
%%%%%%%%%%%%%%%%%%%%%%%%%%%%%%%%%%%%%%%%%%%%%%%%%%%%%
\section{Topological phases at lower fillings}
\label{sec:lower-fillings}

%%%%%%%%%%%%%%%%%%%%%%%%%%%%%%%%%%%%%%%%%%%%%%
\begin{figure*}[ht]
	\begin{center}
		\includegraphics{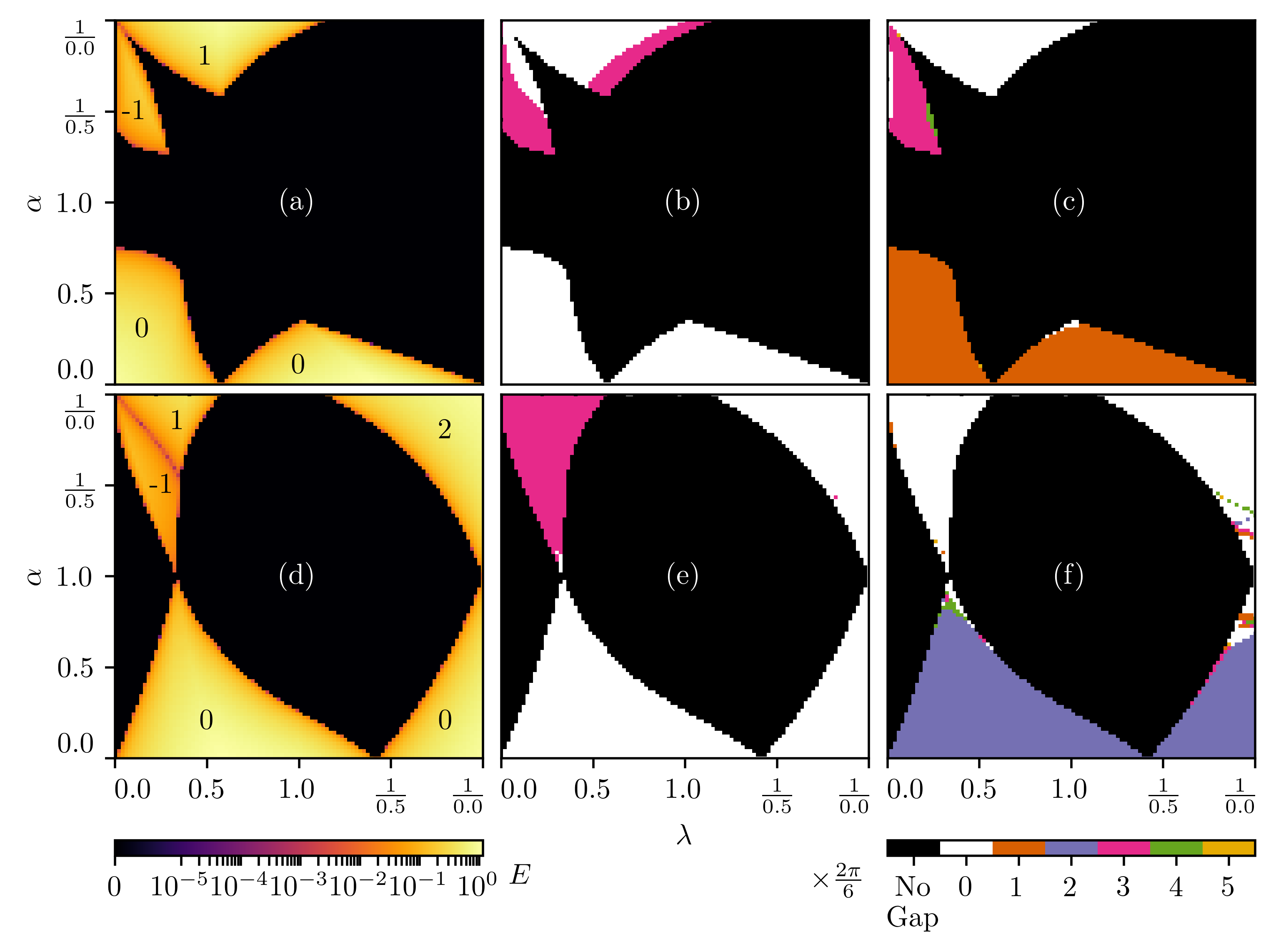}
		\caption{
			(a,d) Energy gap vs.\ $\alpha$ and $\lambda$, (b,e) the $\mathbb{Z}_2$ twist Berry phase vs.\ $\alpha$ and $\lambda$, (c,f) $\mathbb{Z}_6$ Berry phase vs.\ $\alpha$ and $\lambda$.
			First row corresponds to 1/6 filling, the second row to 1/3 filling. Energy gap calculations are performed in momentum space, while $\mathbb{Z}_Q$ twist Berry phases in position space with 16 $\times$ 16 hexamer unit cells.
		}
		\label{fig:lower_filling_diagrams}
	\end{center}
\end{figure*}
%%%%%%%%%%%%%%%%%%%%%%%%%%%%%%%%%%%%%%%%%%%%%%

In the following, we focus on lower fillings which have been less studied in the past. We note that due to the particle-hole symmetry of the spectrum, we expect our findings to hold for the corresponding higher fillings as well. Fig\,\ref{fig:lower_filling_diagrams}\,(a) and (d) show the energy gap at 1/6 and 1/3 fillings for all values of $\alpha$ and $\lambda$. Several gapped phases can be observed, in addition to an extended metallic region.

\subsection{Third Filling}
\label{sec:third-filling}

Also note that the gap is closed for any value of $\alpha$ when $\lambda = 0$. From Fig.\,\ref{fig:lower_filling_diagrams} (e), we can also see that the two left top phases both also have a $\mathbb{Z}_2$ twist Berry phase of $\pi$.

The lower two phases both have a Chern number of zero, $\mathbb{Z}_2$ twist Berry phase of zero, and a $\mathbb{Z}_6$ Berry phase of $2 \times 2 \pi / 6 = 2 \pi / 3$. This matches the analytic result, which is that the $\mathbb{Z}_6$ twist Berry phase is equal to $n \times 2 \pi$ in a phase is adiabatically connected to the decoupled hexamer limit, where $n$ is the filling factor\,\cite{araki2020}.

A Chern number of zero and a non-zero $\mathbb{Z}_6$ Berry phase lead us to look for isolated corner states in the third-filling gap. The pentamer corner structures which gave corner states at $E=0$ across the whole of half-filling phases II and III also give corner states in the third-filling gap at $\alpha = 0$, \ie in the disconnected hexamer cluster limit. This is not true for other corner structures, \ie  monomers, dimers, trimers, or tetramers. We thus have identified the structure with ``pentamer corners'' as the one to observe corner states of the HOTI phases at lower fillings.

Spectra in Fig.\,\ref{fig:slice_spectra_combo} in the Appendix show that the pentamer corner states persist while the gap remains open.  However, it should be noted that corner states can enter a bulk band while the third filling gap is still open, like in Fig.\,\ref{fig:slice_spectra_combo}\,(e) and (g). This is a problem unique to fillings that are not half filling for this system, as a particle-hole symmetry in the Hamiltonian ensures that bands come in positive-negative pairs, and corner states exist at zero energy, so for the corner state to enter the bulk band, the half filling gap must close. A similar behaviour is also seen in the breathing kagome model in a triangular shape~\cite{ezawa2018-kagome}.

\subsection{Sixth Filling}
\label{sec:sixth-filling}

As can be seen in Fig.\,\ref{fig:lower_filling_diagrams} (a), there are 4 unique insulating phases at sixth filling. The two phases at the top of the diagram (for $\alpha > 1$) are again both Chern phases. The phase on the top left has a Chern number of -1, as demonstrated in Figs.\,\ref{fig:hoti_phases} and\,\ref{fig:hoti_phases} (a) and (b), and the phase on the top right has Chern number of +1, see Fig.\,\ref{fig:Chern_phases}\,(c) and (d). Like for third filling, the sixth filling bulk band gap is closed for $\lambda = 0$ (not visible in Fig.\,\ref{fig:lower_filling_diagrams}\,(a)), so neither of these phases are adiabatically connected to the dimer decoupled cluster limit. Portions of these phases have a $\mathbb{Z}_2$ twist Berry phase equal to $\pi$, and where the transition happens within a phases, the twist Hamiltonian has closed at $\theta = \pi$. From Fig.\,\ref{fig:lower_filling_diagrams} (c), we see that the left top phase also has a $\mathbb{Z}_6$ twist Berry phase equal to $\pi$. Note that the portions of the phase with other twist Berry phase values correspond to regions with very small gaps, so are not equal to $\pi$ due to finite size effects. See the discussion for further comments about the interplay between first- and higher-order topology in Sec.\,\ref{sec:discussion}.

The two lower phases both have a Chern number of zero, see Fig.\,\ref{fig:hoti_phases}\,(b) and (d) for example spectra and Berry curvature calculations for the bottom left phase.  As with the two lower phases for third filling, the expected $\mathbb{Z}_6$ twist Berry phase of $\frac{1}{6} \times 2 \pi = \pi / 3$ is seen in the numerical calculations, and persists over the entire phase. Pentamer corner states also are present in the bulk band gap across the phase, as seen in Figs.\,\ref{fig:real_space_geometry} and \ref{fig:slice_spectra_combo}.

%%%%%%%%%%%%%%%%%%%%%%%%%%%%%%%%%%%%%%%%%%%%%%%%%%%%%%%%%%%%%%%%%%%%%%%%%%%
%%%%%%%%%%%%%%%%%%%%%%%%%%%%%%%%%%%%%%%%%%%%%%%%%%%%%%%%%%%%%%%%%%%%%%%%%%%
%%%%%%%%%%%%%%%%%%%%%%%%%%%%%%%%%%%%%%%%%%%%%%%%%%%%%%%%%%%%%%%%%%%%%%%%%%%

%%%%%%%%%%%%%%%%%%%%%%%%%%%%%%%%%%%%%%%%%%%%%%%%%%%%%
%
%                                                      D I S C U S S I O N
%
%%%%%%%%%%%%%%%%%%%%%%%%%%%%%%%%%%%%%%%%%%%%%%%%%%%%%
\section{Discussion}
\label{sec:discussion}

\subsection{Phase boundaries and changes of Chern number}
\label{sec:phase-boundaries}

The Chern phases of this model at half filling are notable for their high Chern numbers. Large changes in Chern number can be observed when crossing phase boundaries, with the largest change occurring from phases VIII to IX. These phases have Chern numbers of -4 and 5, respectively, meaning that the Chern number changes by $\Delta C=9$ moving between these phases. The changes in Chern number can be understood by examining the gap closings. Dirac cones are the sources and sinks of the topological invariant\,\cite{bernevig2013}, where the change in Chern number when crossing a phase boundary is usually equal to the number of Dirac cones that form at the gap closing.
The change in the Chern number by $\Delta C=9$ occurs at the intersection of two phase boundary lines, with one phase boundary line corresponding to 3 Dirac cones at the three $M$ points, and the other line corresponding to the 2 points symmetrically between each of the 3 unique $M$ and $K$/$K'$ points, which gives 6 unique gap closing points on this phase boundary line. 
When crossing over the intersection of these two phase boundary lines, the two gap closings happen at once, which explains the jump by $\Delta C=9$ of the Chern number.  

Aside from the $M$ points and the two points around each of the $M$ points, gap closings can also happen at the $\Gamma$ point, or the $K$ and $K'$ points (as mentioned previously). These gap closings represent the creation of 3, 6, 1 and 2 uniquely placed Dirac cones in the Brillouin zone, respectively.  All the boundaries between phases fall into one of these categories; for example, towards the left of the phase boundary diagram Fig.\,\ref{fig:phase_boundary_diagram}, the phase boundary between the HOTI phases with Chern number 0 and the Haldane phases with Chern number 1 occurs where the gap closes at $\Gamma$ with one Dirac cone. 

%%%
%%%
%%%
\subsection{Double and triple phase boundary points}
\label{sec:sub-ZQ}

There are a few higher-order phase boundary points across the phase boundary diagram. A few of these points are where phase boundary lines cross over each other, \ie a double phase boundary point, similar to what we discussed above for the case $\Delta C=9$. Another higher-order point is at $\alpha = 0, \lambda = 1/\sqrt{3}$, where three lines come out of the one point, \ie a triple phase boundary point. As this point is in the decoupled hexamer limit, the energies are obtained easily through diagonalisation of a $\vec k$ independent $6\times 6$ matrix, and the bands are therefore  flat. Thus the gap closing occurs everywhere in the Brillouin zone (touching of flat bands), rather than just at specific points. 

The corresponding gap closing point does not happen in the opposite limit $\lambda = 1/\sqrt{3}, \alpha = 1/0$ (\ie $\beta=0$), as the dimers are not decoupled for finite $\lambda$, so the bands are $\vec k$ dependent. The point $\alpha = 1/0, \lambda = 1/\sqrt{3}$ is a gap closing point, however, just at the $K/K'$ points. Another higher-order phase boundary point is $\lambda = 1/\sqrt{3}, \alpha = 6$, which is connected to three phase boundary lines (triple phase boundary point). Approaching along $\lambda < 1/\sqrt{3}$, the phase boundary line occurs at $M$, and for $\lambda > 1/\sqrt{3}$, the phase boundary splits into two -- with one phase boundary continuing at $M$, and the other for the two points around $M$ along the $K$ to $K'$ line. This higher-order point is the origin of this later line, which is where the zero energy points bifurcate from $M$, moving closer to $K$ with increasing $\lambda$.

%%%
%%%
%%%
\subsection{Structure of Chern phases}
\label{sec:struc-chern}

%%%%%%%%%%%%%%%%%%%%%%%%%%%%%%%%%%%%%%%%%%%%%%
\begin{figure*}[ht]
\centering
\includegraphics[width = 0.45\textwidth]{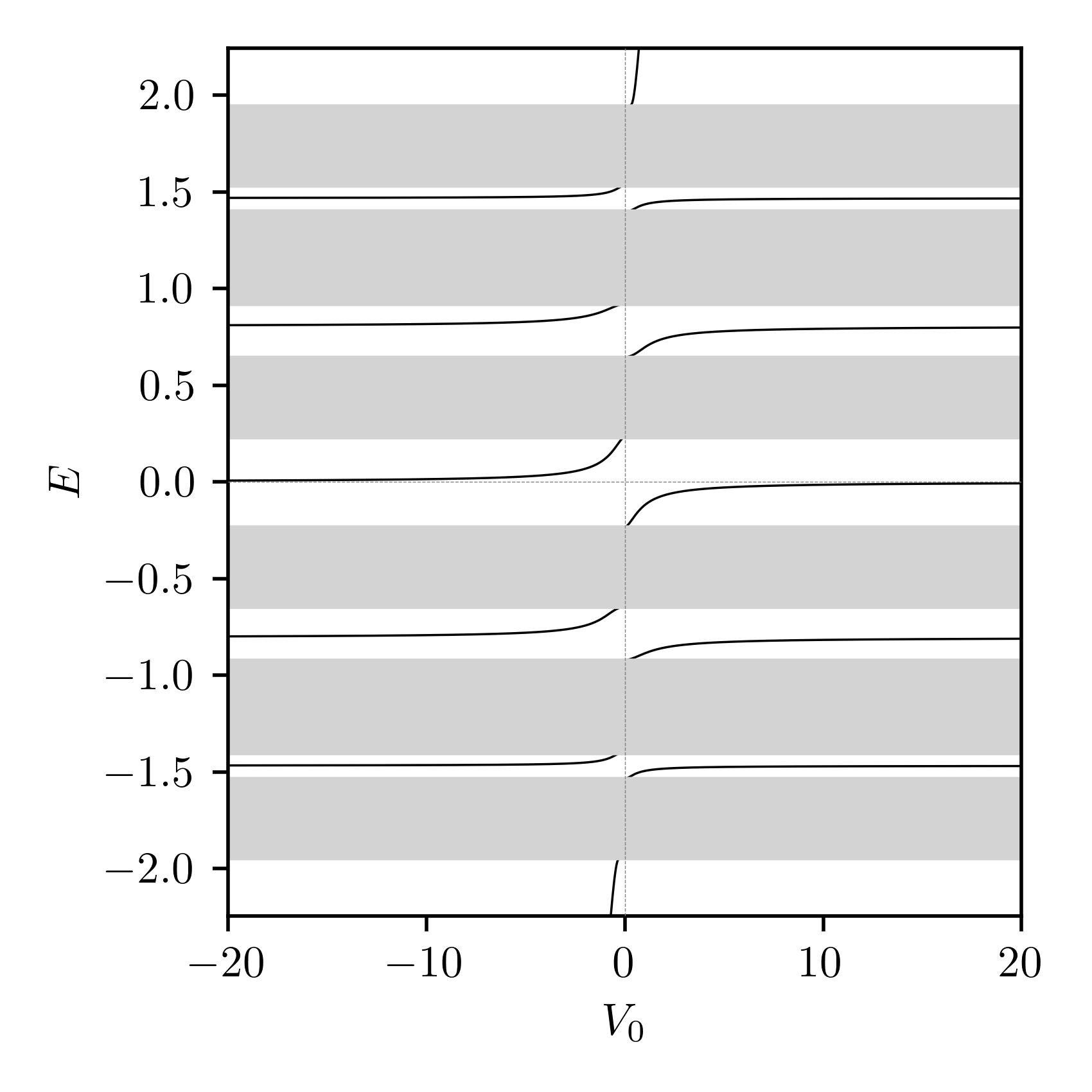}
\includegraphics[width = 0.45\textwidth]{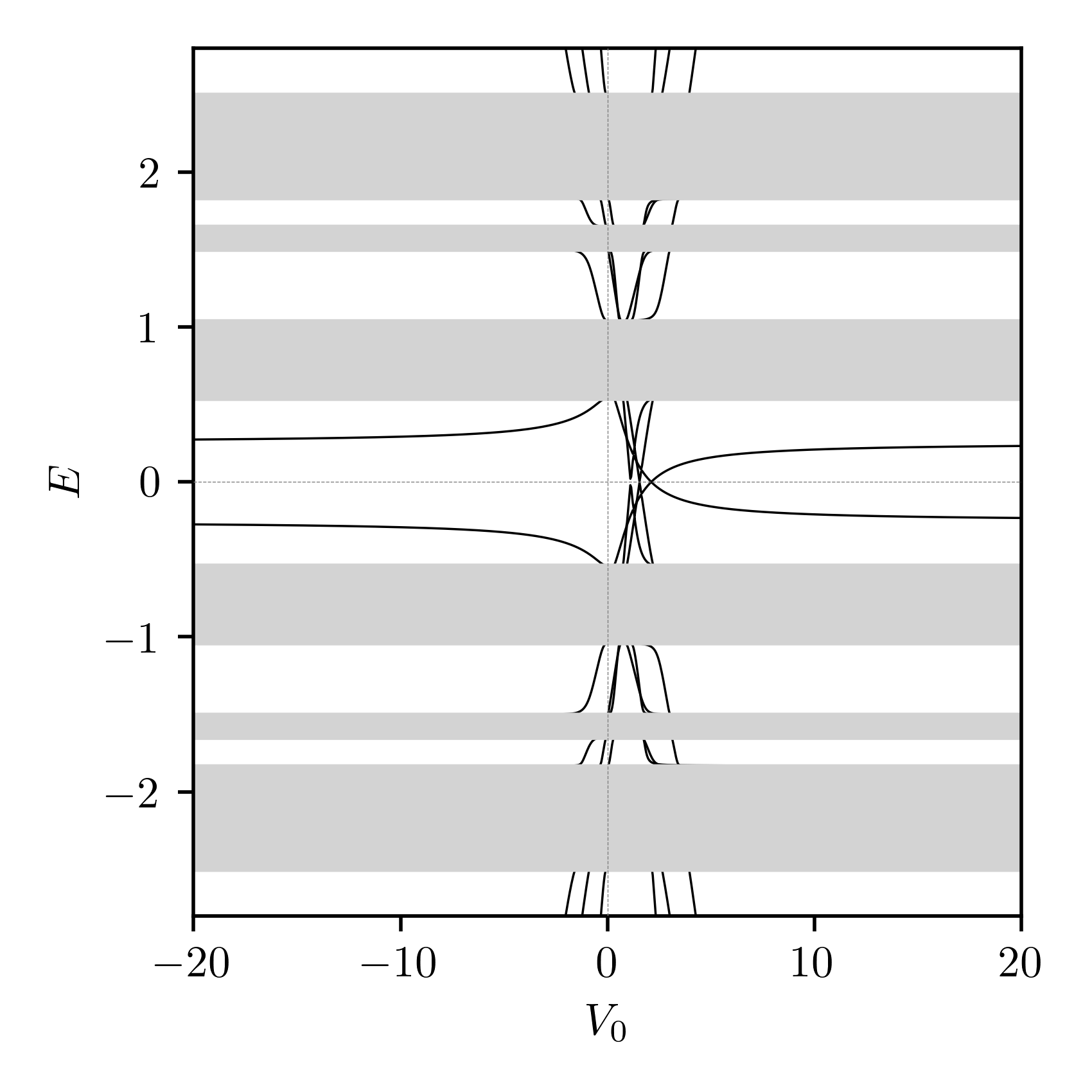}
\caption{Spectral response to an impurity, plotting the spectrum $E$ vs.\ impurity strength $V_0$. Grey are bulk bands, \ie bands in the absence of the impurity. (Left) Site impurity for $\alpha = 1/0.2, \lambda = 0.3$. Each of the gaps exhibits a Chern number of 1. (Right) Hexamer impurity for $\alpha = 0.5, \lambda = 0.3$. The half filling gap exhibits a Chern number 1, while the other gaps have no Chern number.}
\label{fig:impurities}
\end{figure*}
%%%%%%%%%%%%%%%%%%%%%%%%%%%%%%%%%%%%%%%%%%%%%%

The $\mathbb{Z}_Q$ Berry phase calculation is straightforward for phases with no Chern number, \ie the HOTI phases. Without competition from the Chern number, the finite $\mathbb{Z}_Q$ Berry phase $= 2 \pi \nu / Q$, where $\nu =$ filling factor, characterizes a phase adiabatically connected to a decoupled cluster limit, either hexamer or dimer\,\cite{araki2020}. In the absence of dispersive (chiral) edge states, we can engineer incomplete cluster unit cell boundaries to get corner states which remain energetically separated from bulk states within a finite gap, even in lower fillings than half filling.

The story is not so clear for phases with a finite Chern number and $\mathbb{Z}_Q$ Berry phase. We have previously identified such phases, see Fig.\,\ref{fig:half_fill_z6} but also Fig.\,\ref{fig:lower_filling_diagrams}. Not only have we found examples with both finite Chern number and finite Berry phase, but also observed changes of the $\mathbb{Z}_Q$ Berry phases within a fully gapped Chern phase. Such changes of the Berry phase happen at gap closings for a finite twist angle (while the bulk gap, \ie the spectral gap for twist angle $\theta=0$, remains finite). Crucially, none of the Chern phases are adiabatically connected to the decoupled cluster limit. Some of these Chern phases have boundary points that are in a decoupled cluster limit, but not the phases themselves. There are two cases: one is the half-filling Chern phase and $\alpha = 0, \lambda = 1/\sqrt{3}$, and the sixth and third filling phases with $\alpha = 1/0, \lambda = 0$. In both these cases, the gap mentioned is closed a that point, so the Chern phases are not adiabatically connected to these phases. 

This means that finite $\mathbb{Z}_Q$ Berry phase does not characterize a connection to a decoupled cluster limit in a Chern phase. This is not surprising, as the unit cell of a decoupled cluster limit can always be defined in a way that there are no bonds between unit cells. Thus the Bloch matrix is $\vec k$-independent and the bands are flat. That implies that phases adiabatically connected to a decoupled cluster limit must have a Chern number of zero. We might also consider that the finite $\mathbb{Z}_Q$ Berry phase in the half-filling Chern phases might indicate that lower filling band gaps are still connected to the decoupled cluster limit, however this is not the case, as the extension of the $\mathbb{Z}_Q = \pi$ phase in the half-filling phase IV does not overlap with the open gap at lower fillings. 
It is also meaningless to discuss the presence of in gap corner states in Chern phases, as the bulk-boundary correspondence of the first-order topology requires edge states to traverse the entire bulk gap. Therefore, isolated corner states are not observable.

We are tempted to claim that $\mathbb{Z}_Q$ Berry phases in phases which are disconnected from the decoupled cluster limit are not necessarily well-defined; clearly further work is required to clarify this situation. We would like to mention that the $\mathbb{Z}_Q$ Berry phase evaluated in Ref.\,\onlinecite{li2020} defined in momentum space is {\it not} equivalent to the Berry phase computed and discussed in this work (thus we have not explored it any further here).

%%%
%%%
%%%
\subsection{Response to impurities}
\label{sec:impurities}

Since the $\mathbb{Z}_Q$ Berry phase calculation is local invariant (in contrast to the Chern number which is a global quantity), there are also parallels to the way a material responds to local impurities. Following the method described in Refs.\,\onlinecite{slager-15prb,diop2020}, we have tested the spectral response to a positive or negative local impurity potential. The prediction\,\cite{slager-15prb,diop2020} is that a topologically non-trivial phase will show an impurity bound state within the spectral gap, no matter how large the impurity potential might be. In contrast, a trivial phase might feature a fine-tuned in-gap state, but for sufficiently large impurity strength it will have moved into the bulk and not appear within the gap. It was further clarified\,\cite{diop2020} that this characterization does only hold for topological phases which are not caused by a modulation of bond strengths. In particular, higher-order topological insulator phases such as the Kekul\'e phases do not share this impurity response. Instead, the impurity response can be fully understood by the decoupled cluster limit\,\cite{diop2020}.

In the following, we discuss the impurity response within the Haldane phase for $\alpha\not= 1$, \ie for Haldane phase with finite (anti-) Kekul\'e distortion. We have tested single site, bond and hexamer impurities. A hexamer impurity consists of six bond impurities sharing the bonds of the same hexagon. Our analysis is consistent with previous findings\,\cite{diop2020}. Everywhere in the Haldane phase, we find one (multiple) in-gap bound state(s) for arbitrary impurity potential strength $V_0$ for a site (hexamer) impurity, see Fig.\,\ref{fig:impurities} at half filling. In this figure, we have exemplarily shown two parameter points within the Haldane phase; the left panels shows the spectral response for a site impurity, while the right panel the one for a hexamer impurity. The parameters for the left panel of Fig.\,\ref{fig:impurities} are chosen such that the lower filling gaps also exhibit a Chern number $C=1$. In contrast, the parameters for the right panel are chosen such that the lower filling gaps possess zero Chern number. As expected, only for a fine-tuned parameter range of $V_0$ in-gap bound states are present, but absent for large positive and negative $V_0$. In our analysis, we cannot observe any features in the spectral response to local impurities which would explain the finite Berry phases within the Chern phases.

%%%%%%%%%%%%%%%%%%%%%%%%%%%%%%%%%%%%%%%%%%%%%%%%%%%%%
%
%                                                      S U M M A R Y
%
%%%%%%%%%%%%%%%%%%%%%%%%%%%%%%%%%%%%%%%%%%%%%%%%%%%%%
\section{Summary}
\label{sec:summ}

We have analyzed the combined effects of Haldane's term leading to a Chern insulator and a Kekul\'e (anti-Kekul\'e) distortion leading to a higher-order topological insulating phase on the honeycomb lattice. Interplay and competition of both terms leads to a surprisingly rich phase diagram with 12 phases at half filling, and several more phases at lower fillings. All phases can be classified by either the Chern number or the $\mathbb{Z}_Q$ Berry phase. In addition, we have presented a thorough investigation of the spectral properties for periodic boundary conditions, ribbon spectra geometry as well as open boundary conditions.
Most interestingly, (i) we have found phases with high Chern numbers, (ii) a novel higher-order topological insulator phase, as well as (iii) a Chern phase corresponding to two coupled Kagome Chern insulators. Furthermore, we have explored the insulating phases at lower fillings, and found again first-order and second-order topological phases. Finally, we have identified real-space structures which feature corner states not only at half but also at third and sixth fillings, in agreement with the quantized $\mathbb{Z}_Q$ Berry phase of the corresponding phases. It remains an exciting problem for future work how the topological phase diagram established in this paper will evolve in the presence of electron-electron interactions\,\cite{rachel2018}.

%%%%%%%%%%%%%%%%%%%%%%%%%%%%%%%%%%%%%%%%%%%%%%%%%%%%%%%%%%%%%%%%%%%%%%%%%%%

\acknowledgments

Discussions with S.\ Samba-Diop and T.\ L.\ Hughes are acknowledged.
This work is funded by the Australian Research Council through Grants No.\ FT180100211 and DP200101118 (S.\ R.).
This work is also supported by JSPS KAKENHI, Grants No. JP17H06138 (T.\ M. and Y.\ H.) and No. JP20K14371 (T.\ M.).
This research was supported in part by the National Science Foundation under Grant No.\ NSF PHY-1748958 (S.\ R.).

%%%%%%%%%%%%%%%%%%%%%%%%%%%%%%%%%%%%%%%%%%%%%%%%%%%%%%%%%%%%%%%%%%%%%%%%%%%

\bibliographystyle{prsty} %abbrvnat.bst} %plainnat.bst} %unsrt}
\bibliography{competingtopologies}

\appendix
\section{Additional information}

%%%%%%%%%%%%%%%%%%%%%%%%%%%%%%%%%%%%%%%%%%%%%%
\begin{figure*}[phtb]
	\centering
	\includegraphics[width=\textwidth]{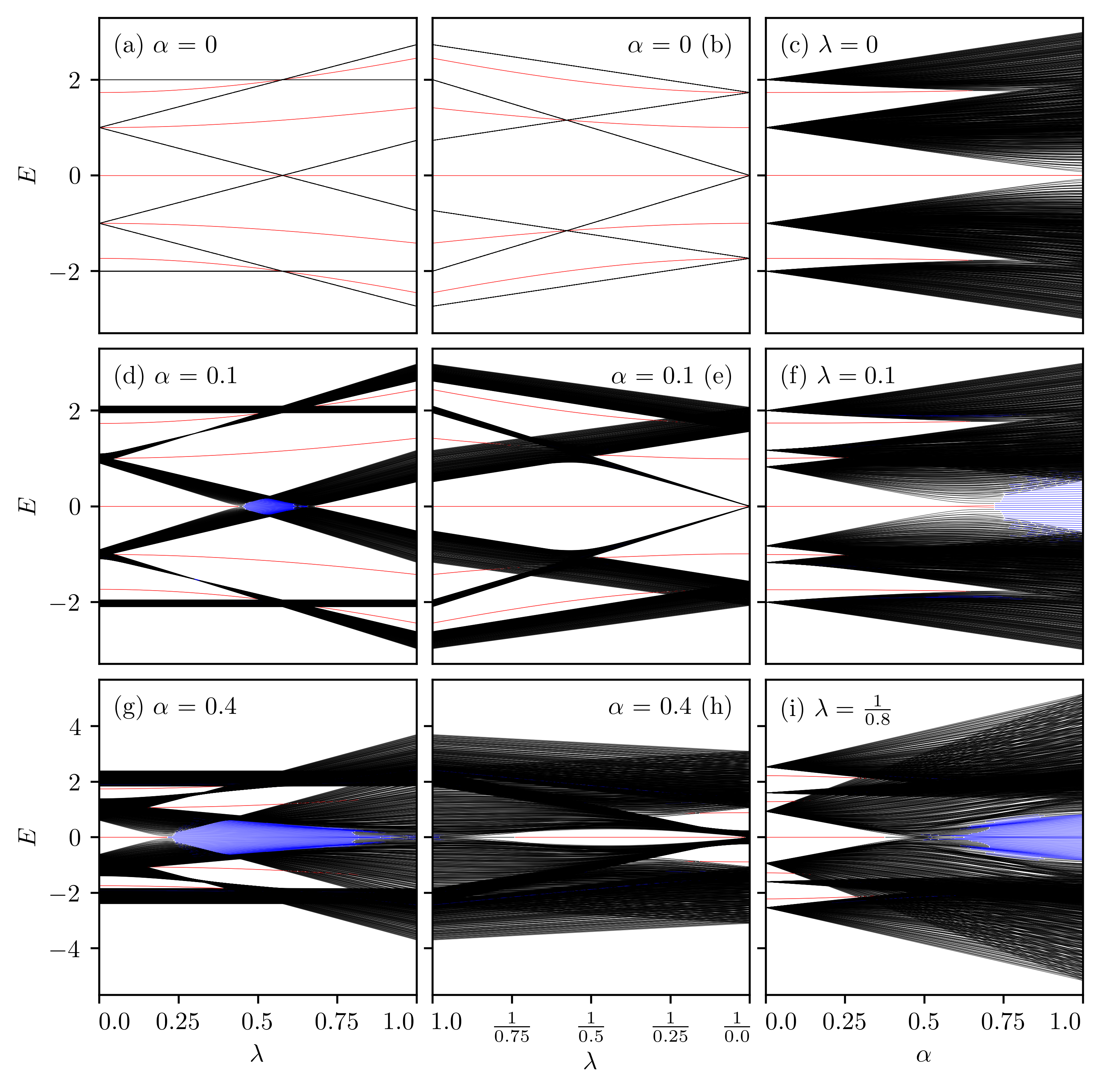}
	\caption{Real space spectra with open boundary conditions, selected vertical or horizontal cuts through the phase diagram Fig.\,\ref{fig:phase_boundary_diagram}. Lattice geometry contains 16 $\times$ 16 hexamer unit cells but with an site from each of the 120$^\circ$ corners removed; the lattice structure corresponds to the smaller 10$\times$10 case shown in Fig.\,\ref{fig:real_space_geometry}. (a,b,d,e,g,h) $\lambda$ cuts for fixed values of $\alpha$, see legends. (c,f,i) $\alpha$ cuts for fixed values of $\lambda$, see legends.
Energies corresponding to eigenstates with $\geq$35\% localised on the 120$^\circ$ corner pentamers are red, and energies corresponding to eigenstates with $\geq$55\% localised on a hexamer unit cell which is on the edge of the lattice (but not a corner state) are blue.}
	\label{fig:slice_spectra_combo}
\end{figure*}
%%%%%%%%%%%%%%%%%%%%%%%%%%%%%%%%%%%%%%%%%%%%%%

%%%%%%%%%%%%%%%%%%%%%%%%%%%%%%%%%%%%%%%%%%%%%%
\begin{figure*}[phtb]
	\centering
	\includegraphics[width=\textwidth]{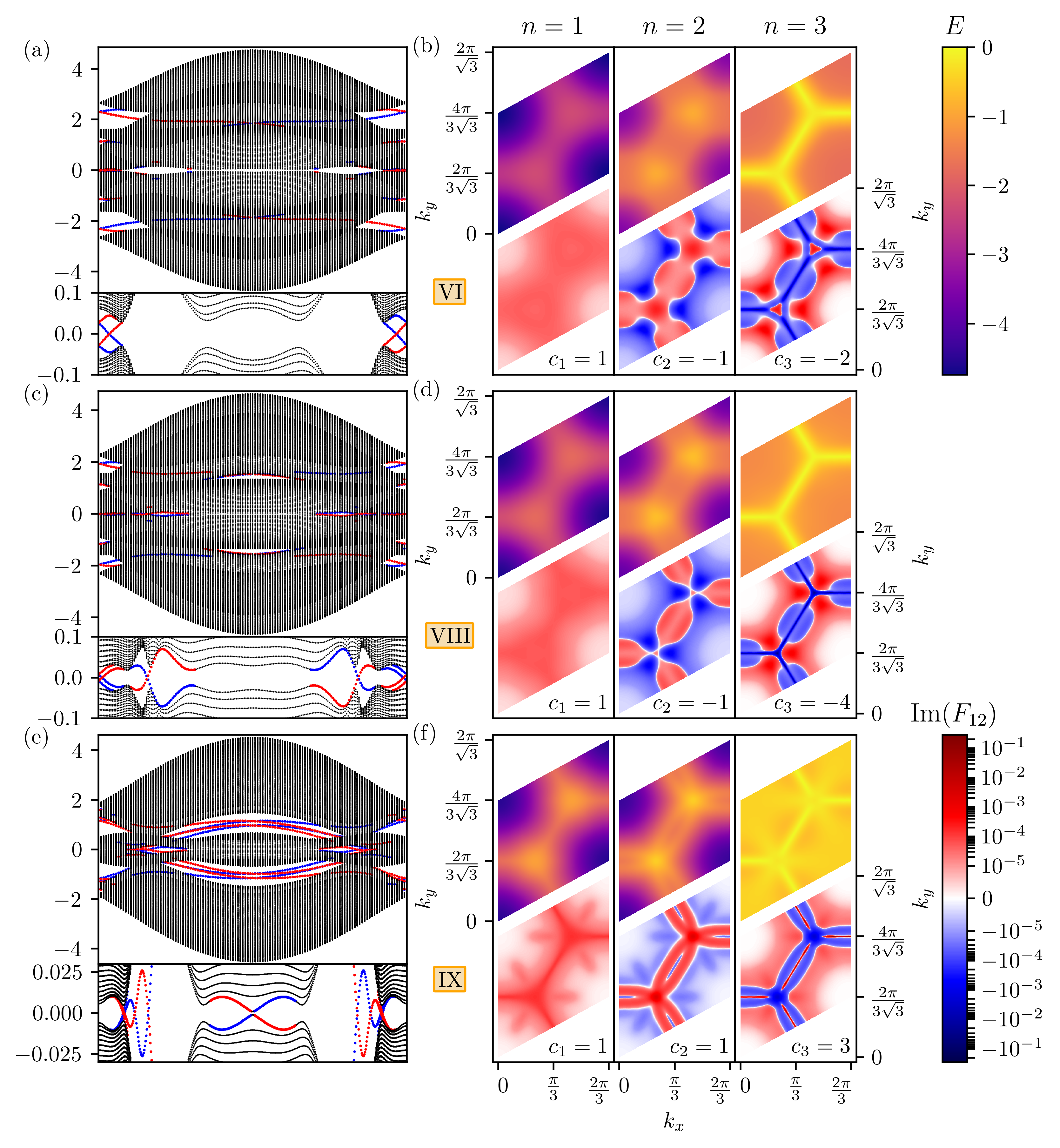}
	\caption{Ribbon spectra (a, c, e), spectra in momentum space (top) and Berry curvature (bottom) over the Brillouin zone (b, d, f). First row (a, b) is phase VI with $C=-2$, parameters used $\alpha = \frac{1}{0.38}, \lambda = 1$; second row (c, d) is phase VIII with $C=-4$, parameters used $\alpha = \frac{1}{0.466}, \lambda = \frac{1}{0.693}$ ; third row (e, f) is phase IX with  $C=5$, parameters used $\alpha = 1/0.5595, \lambda = 1/0.2$. For the ribbon spectra, blue (red) dots are states with $\geq$75\% of the wavefunction contained on the left (right) of the ribbon. Ribbon spectra are calculated for a width of 100 unit cells (600 atomic sites), with a $k$ resolution of $2 \pi / 125$ (200 unit cells with a $k$ resolution of $2 \pi / 200$ for the zoomed in regions in (a,c); 800 unit cells with a $k$ resolution $2 \pi / 500$ for zoomed-in regions in (e).	Note that in (c), there are two closely spaced but separate edge states traversing the bulk gap at $E=0$ on each side, and one in the middle. 
	Momentum space spectra and Berry curvature evaluated on a $500 \times 500$ grid.}
	\label{fig:appendix_phases_1}
\end{figure*}
%%%%%%%%%%%%%%%%%%%%%%%%%%%%%%%%%%%%%%%%%%%%%%

%%%%%%%%%%%%%%%%%%%%%%%%%%%%%%%%%%%%%%%%%%%%%%
\begin{figure*}[phtb]
	\centering
	\includegraphics[width=\textwidth]{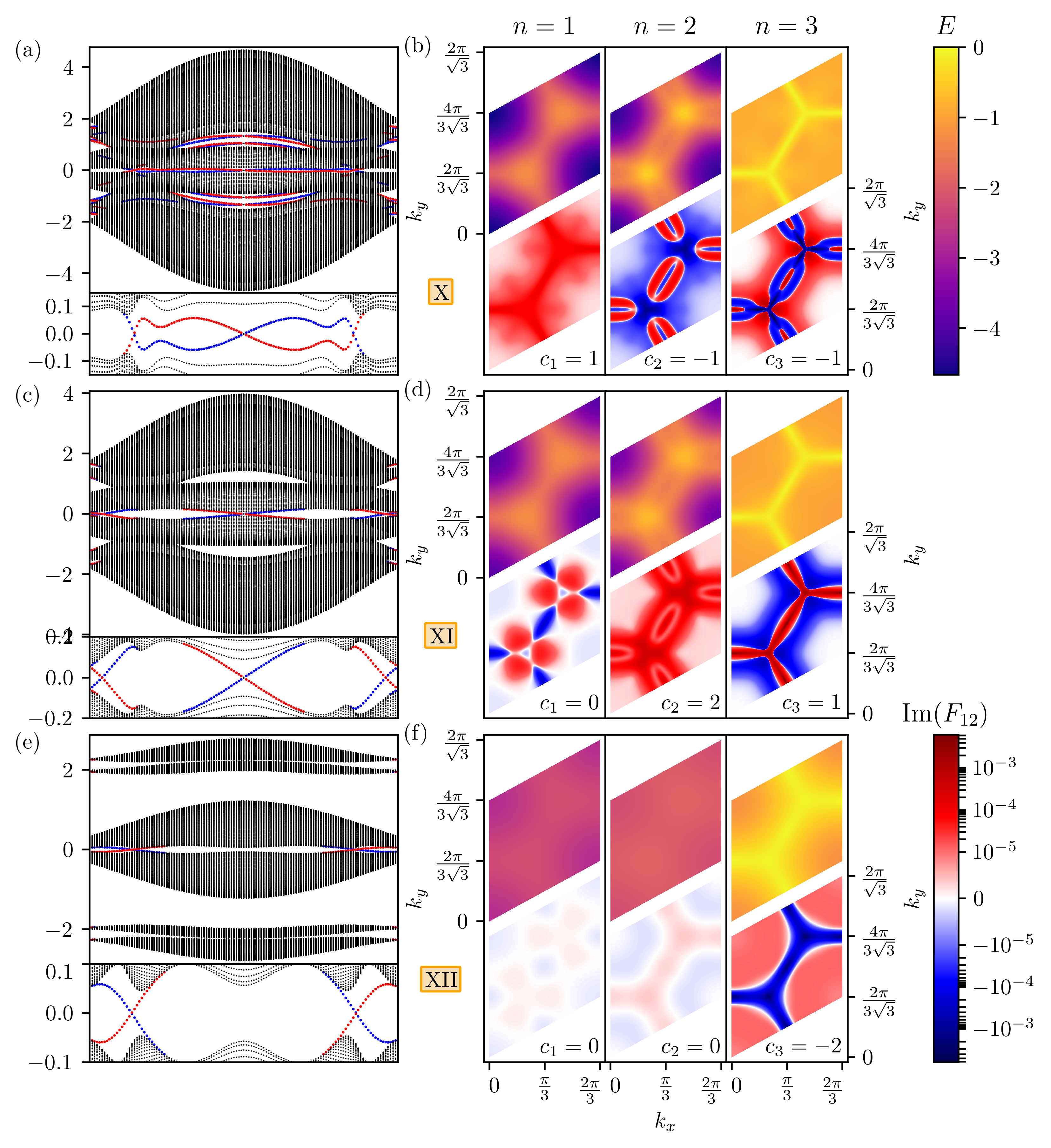}
	\caption{Ribbon spectra (a, c, e), spectra in momentum space (top) and Berry curvature (bottom) over the Brillouin zone (b, d, f). First row (a, b) is phase X with $C=-1$, parameters used $\alpha = \frac{1}{0.6}, \lambda = \frac{1}{0.4}$; second row (c, d) is phase XI with $C=3$, parameters used $\alpha = 0.6, \lambda = \frac{1}{0.4}$; third row (e, f) is phase XII with $C=-2$, parameters used $\alpha = 0.22, \lambda = 0.8$. Displayed below Berry curvature plots are the Chern numbers $c_n$ of each band. For the ribbon spectra, blue (red) dots are states with $\geq$75\% of the wavefunction contained on the left (right) of the ribbon. Ribbon spectra are calculated for a width of 100 unit cells (600 atomic sites), with a $k$ resolution of $2 \pi / 125$. Momentum space spectra and Berry curvature evaluated on a $500 \times 500$ grid.}
	\label{fig:appendix_phases_2}
\end{figure*}
%%%%%%%%%%%%%%%%%%%%%%%%%%%%%%%%%%%%%%%%%%%%%%

\end{document}